\documentclass{aa}

\usepackage[varg]{txfonts}
\usepackage{graphicx}
\usepackage{subcaption}
\usepackage{amsmath}
\usepackage{mathrsfs}
\usepackage{float}
\usepackage{multirow}
\usepackage{booktabs}
\usepackage{tabularx}
\usepackage{makecell}
\usepackage{siunitx}
\usepackage{natbib}
\usepackage{hyperref}
\usepackage{xcolor}
\usepackage{caption}
\DeclareUnicodeCharacter{0327}{\c}
\makeatletter
\let\linenumbers\relax
\let\nolinenumbers\relax
\makeatother

\begin{document}

\title{Effects of Stellar X-ray Photoevaporation on Planetesimal Formation via the Streaming Instability}
%\subtitle{}

\author{Xuchu Ying\inst{1}
        \and 
        Beibei Liu\inst{1,}\thanks{Corresponding author: bbliu@zju.edu.cn}
        \and
        Haifeng Yang\inst{1}
        \and 
        Joanna Dr{\k a}{\. z}kowska\inst{2}
        \and
        Sebastian M. Stammler\inst{3}
        \and
        Zhaohuan Zhu\inst{4}
        \and \\
        Linn E.J. Eriksson\inst{5}
        \and
        Hongping Deng\inst{6}
        \and
        Bin Liu\inst{1}
        \and
        Ping Chen\inst{1}
        }

\institute{
    Institute for Astronomy, School of Physics, Zhejiang University, Hangzhou 310058, China\\
    \email{xcying@zju.edu.cn},
    \and
    Max Planck Institute for Solar System Research, Justus-von-Liebig-Weg 3, 37077 Göttingen, Germany
    \and
    University Observatory, Faculty of Physics, Ludwig-Maximilians-Universität München, Scheinerstr. 1, 81679 Munich, Germany
    \and
    Department of Physics and Astronomy, University of Nevada, 4505 South Maryland Parkway, Las Vegas, NV 89154, USA
    \and
    Department of Astrophysics, American Museum of Natural History, 200 Central Park West, New York, NY 10024, USA
    \and
    Shanghai Astronomical Observatory, Chinese Academy of Sciences, Nandan Rd 80th, Shanghai 200030, China
    }

\date{xx.xx.xx; yy.yy.yy}

\abstract
% context
{The formation of planetesimals via the streaming instability (SI) is a crucial step in planet formation, yet its triggering conditions and efficiency are highly sensitive to both disk properties and specific evolutionary processes.}
% aims
{We aim to study the planetesimal formation via the SI, driven by the stellar X-ray photoevaporation during the late stages of disk dispersal, and quantify its dependence on key disk and stellar parameters.}
% methods
{We use the \texttt{DustPy} code to simulate the dust dynamics including coagulation, fragmentation, and radial drift in a viscously accreting disk undergoing stellar X-ray photoevaporation.}
% results
{Stellar X-rays drive the disk dispersal, opening a cavity at a few au orbital distance and inducing the formation of an associated local pressure maximum. This pressure maximum acts as a trap for radially drifting dust, therefore enhancing the dust density to the critical level required to initiate the streaming instability and the subsequent collapse into planetesimals. 
The fiducial model produces $31.4 \ M_{\oplus}$ of planetesimals with an initial dust to final planetesimal conversion efficiency of $20.4\%$. This pathway is most efficient in larger disks with higher metallicities, lower viscosities, higher dust fragmentation threshold velocities, and/or around stars with higher X-ray luminosities.}
% conclusions
{This work demonstrates that stellar X-ray photoevaporation is a robust and feasible mechanism for triggering planetesimal formation via the SI during the final clearing phase of protoplanetary disk evolution.}


\maketitle

\section{Introduction}
\label{Sect.1}
Planet formation remains one of the most challenging problems in modern astrophysics
\citep{Safronov1969,Lissauer1993,Pollack1996}.
Over the past three decades, exoplanet surveys using radial velocity, transit, microlensing, and direct imaging techniques have demonstrated that planetary systems are a common outcome of star formation, with the population of detected planets spanning a wide range of masses and orbital architectures \citep{mayor2011harps,Batalha2013,Cassan2012,Mulders2018,zhu2021exoplanet}.
The diversity of these systems---from compact super-Earth chains to distant gas giants---provides crucial constraints on the physics of the protoplanetary disk evolution and planet formation
(see reviews of \citealt{Liu&Ji2020,drazkowska2023planet}).

The core accretion scenario offers a broadly successful framework for linking these observations to the final planetary architectures \citep{Pollack1996,Ida&Lin2004,Mordasini2012,Bitsch2015,liu2019super,Liu&Ji2020,emsenhuber2021new,pan2025dependence}. 
In this picture, micron-sized dust particles coagulate into mm-cm pebbles
\citep{Blum&Wurm2008,birnstiel2012simple,pinilla2012trapping,Birnstiel2016},
but further growth is uncertain: laboratory experiments \citep{blum1993experimental,blum2000experiments,Blum&Wurm2008,Guttler2010}, and numerical simulations \citep{Brauer2008,Zsom2010,Okuzumi2012,Kataoka2013,drkazkowska2016} all indicate that bouncing, fragmentation, and rapid inward drift prevent pebbles from directly forming kilometer-sized planetesimals. 
These barriers are most severe inside ice lines or in disks with low dust-to-gas ratios, where pebbles drift inward before significant growth can occur
\citep{drazkowska2023planet}. 
The conversion of mm-sized solids into planetesimals therefore remains one of the least understood stages of planet formation
\citep{johansen2014multifaceted,Liu&Ji2020,drazkowska2023planet}, demanding an additional, efficient concentration mechanism \citep{Birnstiel2016,drkazkowska2016}.

The streaming instability (SI) is widely considered a leading candidate for overcoming these barriers \citep{youdin2005streaming,Johansen2007,bai2010dynamics,carrera2015form,li2021thresholds,Lim_2024,lim2025streaming,lim2025probing}. 
By tapping the relative motion between gas and solids, the SI concentrates drifting solids into dense filaments that can collapse gravitationally into planetesimals. 
Its activation, however, depends sensitively on local dust-to-gas ratio, particles' Stokes number, and disk turbulence \citep{bai2010dynamics,yang2017concentrating,Drazkowska&Dullemond2014,Lim_2024}, and local dust-to-gas ratio several times that of the sun is typically required \citep{Johansen2009,carrera2015form,drkazkowska2016,Liu&Ji2020}.
This raises a key question: how and where can protoplanetary disks enhance the dust-to-gas ratio sufficiently to trigger the SI?

High-resolution imaging of protoplanetary disks has revealed ubiquitous substructures---such as rings, gaps, and cavities---which are interpreted as evidences that dust growth and radial drift are already actively reshaping disks at very young ages \citep{Huang&Andrews2018,Dullemond2018,andrews2020observations}.
Pressure bumps are thought to be natural sites for halting dust drift and locally boosting the metallicity. 
They may arise from gap opening by massive planets
\citep{Kobayashi2012,zhu2012dust,gonzalez2015accumulation,eriksson2020pebble}, at the inner edges of magnetically inactive dead zones \citep{varniere2006reviving,Lyra2009,Dzyurkevich2010}, from zonal flows generated by hydrodynamical or magnetic instabilities
\citep{johansen2009zonal,Dittrich2013,bai2014hall,Raettig2015,xu&bai2022turbulent,lehmann2025convective}, in the ice line regions \citep{ida2016formation,drkazkowska2017planetesimal,schoonenberg2017planetesimal}, or at the outer edges of photoevaporative cavities \citep{carrera2017planetesimal,ercolano2017x}.

High-energy radiation, particularly X-rays from young stellar hosts, plays a critical role in clearing gas from protoplanetary disks during their late evolutionary stages \citep{owen2010radiation,Owen2012,ercolano2017x,wang2017wind,Jennings2018,Picogna2019}. This radiation heats and ionizes the upper layers of the disk, driving the photoevaporative winds that steadily deplete the gas. This process initially opens a cavity at a few au orbital distance for a solar-mass star (closer-in for a lower-mass star). 
As depletion continues, the cavity expands outward \citep{owen2010radiation,alexander2014dispersal,ercolano2017dispersal,Picogna2019}. A local pressure maximum forms near the expanding cavity's edge, which efficiently traps drifting dust \citep{pinilla2018homogeneous,garate2021large,garate2023millimeter}. This evolving pressure bump can enhance the local metallicity and trigger the SI, potentially facilitating late-stage planetesimal formation \citep{carrera2017planetesimal,ercolano2017x}. Similarly, \cite{throop2005can} pointed out that the external photoevaporation from the nearby massive stars removes the gas efficiently, while the dust population remains largely unaffected. This can also increase the disk dust-to-gas ratio.

This paper investigates whether stellar X-ray photoevaporation can produce the conditions required to trigger the SI and form planetesimals. We employ the \texttt{DustPy} code to couple models of viscous gas evolution with dust coagulation, fragmentation, and radial drift, explicitly including photoevaporation to simulate the formation and inside-out movement of a disk cavity with its associated pressure bumps. By applying established SI criteria to these evolving disks, we determine the circumstances under which the photoevaporation-driven dispersal promotes the planetesimal formation.

Our paper is organized as follows. The model is introduced in Sect.~\ref{Sect.2}. 
Our main results and parameter exploration are presented in Sect.~\ref{Sect.3}. 
The implications and uncertainties are discussed in Sect.~\ref{Sect.4}. The conclusions are drawn in Sect.~\ref{Sect.5}. 

\section{Method}
\label{Sect.2}
In this section, we introduce the basics of our model. The prescriptions of disk evolution, stellar X-ray photoevaporation, and planetesimal formation are described in Sect.~\ref{Sect.2.1}, ~\ref{Sect.2.2} and Sect.~\ref{Sect.2.3}, respectively, whereas the simulation setup is given in Sect.~\ref{Sect.2.4}.

\subsection{Disk model}
\label{Sect.2.1}
\subsubsection{Gas component}
    The initial gas disk surface density is adopted from the self-similar solution of \cite{lynden1974}
    \begin{equation}
        \Sigma_{\rm g0} = \frac{M_{\rm disk}}{2\pi R_{\rm c}^2}
                        \left(\frac{r}{R_{\rm c}}\right)^{-1} {\rm exp}\left(-\frac{r}{R_{\rm c}}\right),
        \label{sigma_g}
    \end{equation}
    where \(M_{\rm disk}\) is the onset gas disk mass, \(R_{\rm c}\) is the disk characteristic radius, and \(r\) is the radial distance from the central star.
    
    Considering a passively irradiated disk with a constant irradiation angle of \(\theta{=}0.05\), the disk midplane temperature can be written as
    \begin{equation}
        T = \left(\frac{\theta\,L_{\star}}{8\pi r^2\,\sigma_{\rm SB}}\right)^{1/4}
        =220\ {\rm K}\,\left(\frac{r}{\rm au}\right)^{-1/2},
        \label{T_mid}
    \end{equation}
    where $L_{\star}{=}4\pi \sigma_{\rm SB} R_{\star}^2 T_{\rm eff}^4$ is the stellar luminosity, and $\sigma_{\rm SB}$ is the Stefan-Boltzmann constant. 
    We choose a premain-sequence solar-mass central star with a stellar radius of $R_\star{=}2R_\odot$ and an effective temperature of $T_{\rm eff}{=}5772$ K.
    
    The aspect ratio of the gas disk is therefore expressed as 
    \begin{equation}
        h_{\rm g} = \frac{c_{\rm s}}{v_{\rm K}} = \frac{H_{\rm g}}{r} = 0.03  \left( \frac{r }{\rm au }\right)^{1/4},
    \end{equation}
    where $c_{\rm s}$ is the isothermal gas sound speed, $v_{\rm K}{=}\sqrt{G M_{\star}/r} $ is the Keplerian velocity, $G$ is the gravitational constant, $M_{\star}$ is the stellar mass, and $H_{\rm g}$ is the gas disk scale height. 

    The gas pressure gradient parameter is defined as 
    \begin{equation}
        \eta = - \frac{h_{\rm g}^{2}}{2}\frac{\partial{\rm ln} P_{\rm g}}{{\partial} {\rm ln} r},
        \label{headwind}
    \end{equation}
    where $P_{\rm g}$ is the midplane gas pressure. 
    Another useful dimensionless parameter that quantifies the gas pressure effect can be defined as
    \begin{equation}
        \Pi = \frac{\eta v_{\rm K}}{c_{\rm s}},
        \label{Pi}
    \end{equation}
    where $\eta v_{\rm K}$ is the gas headwind velocity, indicating how much the azimuthal velocity of the gas deviates from the local Keplerian velocity.

    The gas surface density is evolved through the viscous advection-diffusion equation \citep{lynden1974},
    \begin{equation}
        \frac{\partial}{\partial t}\Sigma_{\rm g} + \frac{1}{r}\frac{\partial}{\partial r}(r\Sigma_{\rm g} v_{\rm g}) = \dot{\Sigma}_{\rm PE,g},
        \label{evo_g}
    \end{equation}
    where \(v_{\rm g}\) is the radial gas velocity and the sink term \(\dot{\Sigma}_{\rm PE,g}\) is the gas surface density decay rate driven by the stellar X-ray radiation, the prescription of which will be presented in Sect.~\ref{Sect.2.2}.
  
    The radial viscous velocity of the gas reads
    \begin{equation}
        v_{\rm g} = -\frac{3}{\Sigma_{\rm g} \sqrt{r}} \frac{\partial}{\partial r}
        (\Sigma_{\rm g} \nu \sqrt{r}),
        \label{v_visc}
    \end{equation}
    where \(\nu{=}\alpha c_{\rm s} H_{\rm g}\) is the gas turbulent viscosity, and \(\alpha\) is the dimensionless viscous parameter \citep{shakura1973black}.

\subsubsection{Dust transport}
   The advection-diffusion equation for the dust is given by \cite{clarke1988diffusion}
    \begin{equation}
        \begin{split}
            \frac{\partial}{\partial t}\Sigma_{{\rm d}}\,+ \,\frac{1}{r}\frac{\partial}{\partial r}
            \Bigg[r\Sigma_{{\rm d}} v_{{\rm d}} &\,-\,
            rD_{\rm d} \Sigma_{\rm g} \frac{\partial}{\partial r}
            \left(\frac{\Sigma_{{\rm d}}}{\Sigma_{\rm g}}\right)\Bigg]
            = \,\dot{\Sigma}_{{\rm PE,d}} + \dot{\Sigma}_{{\rm plt}},
        \end{split}
        \label{evo_d}
    \end{equation}
    where \(\Sigma_{{\rm d}}\) and \(v_{{\rm d}}\) are the surface density and radial velocity of dust, \(\dot{\Sigma}_{{\rm PE,d}}\) and \(\dot{\Sigma}_{{\rm plt}}\) are the dust mass-loss rates driven by the stellar X-ray photoevaporation and planetesimal formation, respectively. Since the dust coagulation is considered, in practical, the above transport equation is solved for dust with different mass bins. 
    The disk metallicity is defined as $Z {=} \Sigma_{\rm d}/\Sigma_{\rm g}$.

    The dust diffusivity taken from \cite{youdin2007particle} reads
    \begin{equation}
        D_{\rm d} = \frac{D_{\rm g}}{1+{\rm St}^2} = \frac{\delta_{\rm r}c_{\rm s} H_{\rm g}}{1+{\rm St}^2},
        \label{diffusion}
    \end{equation}
    where $D_{\rm g}{=}\delta_{\rm r}c_{\rm s} H_{\rm g}$ is the gas diffusivity and the dimensionless turbulent parameter \(\delta_{\rm r}\) represents the turbulent diffusion in the radial direction. We set \(D_{\rm d}{=}0\) for most of our simulations. Including the dust diffusivity has a negligible effect on our results, since the timescale for dust to diffuse across its trapping region is much longer than that of planetesimal formation (see discussions in Sect. \ref{Sect.4.2}).
        
    The aerodynamic size of the dust particles can be quantified by their Stokes number \citep{whipple1972certain,whipple1973radial,weidenschilling1977aerodynamics},
    \begin{equation}
        {\rm St} = \begin{cases}
            \frac{\pi}{2}\frac{a_{\rm d}\rho_{\bullet}}{\Sigma_{\rm g}}& 
            \ \text{at}\ \ a_{\rm d} \leq \frac{9}{4}\lambda_{\rm mfp},\\
            \\
            \frac{2\pi}{9}\frac{a_{\rm d }^2\rho_{\bullet}}{\lambda_{\rm mfp} \Sigma_{\rm g}}&
            \ \text{at}\ \ a_{\rm d} > \frac{9}{4}\lambda_{\rm mfp}.\\
        \end{cases}
    \end{equation}
    Here \(a_{\rm d}\) is the particle's physical size, \(\lambda_{\rm mfp}\) is the mean free path of disk gas, and \(\rho_{\bullet}\) is the particle's internal density, chosen to be  $1.67\ {\rm g\,cm^{-3}}$ as the default value. 
    
    The radial dust velocity is given by
    \begin{equation}
        v_{{\rm d}}=\frac{v_{\rm g}+2v_{\rm drift,max}{\rm St}
        }{1 + {\rm St}^2},
        \label{v_d}
    \end{equation}
    with the maximum drift velocity $v_{\rm drift,max} {=}{-} \eta v_{\rm K}$.
  
\subsubsection{Dust coagulation}
    In our simulations, the dust coagulation is calculated by solving the Smoluchowski equation \citep{smoluchowski1916drei,Stammler_2022}. Basically, the dust initially grows through pair-wise collisions until reaching the radial drift barrier or the fragmentation barrier.

    Although our simulations employ the \texttt{DustPy} code \citep{Stammler_2022}, which directly solves the full Smoluchowski equation, we briefly introduce the concepts of the fragmentation and drift barriers to provide physical intuition for the dust growth process. These barriers describe the limiting regimes in which dust growth is halted by either fragmentation or radial drift, and they are commonly used to interpret the size limits of dust in protoplanetary disks.

    The dust growth, fragmentation, and radial drift have been investigated in-depth  \citep{Safronov1969,weidenschilling1977aerodynamics,weidenschilling1980dust,nakagawa1986settling,dominik1997physics,Brauer2008,Blum&Wurm2008,birnstiel2012simple}.
    Assuming that the relative velocity between dust is driven by disk turbulence \(v_{\rm rel} {\simeq} \sqrt{3 \delta_{\rm t} \rm St}\,c_{\rm s}\) \citep{ormel2007closed}, the growth timescale taken from \cite{birnstiel2012simple} reads
    \begin{equation}
        \tau_{\rm growth}=\frac{a_{\rm d}}{\dot{a}_{\rm d}}
        =\frac{\rho_\bullet a_{\rm d}}{\rho_{\rm d}v_{\rm rel}}
        \simeq \frac{1}{Z\Omega_{\rm K}},
        \label{tau_grow}
    \end{equation}
    where $\rho_{\rm d}$ is the dust volume density, and $\Omega_{\rm K}{=}v_{\rm K}/r$ is the Keplerian frequency.
    The dust radial drift timescale adopted from \cite{birnstiel2012simple} reads
    \begin{equation}
        \tau_{\rm drift}=\frac{r}{v_{\rm d}} \simeq \frac{1}{ \eta {\rm St}\Omega_{\rm K}}.
    \end{equation}
    When the drift timescale is shorter than the growth timescale, the in-situ growth cannot proceed. By equating these two timescales, one can derive the particles' Stokes number in the radial drift barrier as 
    \begin{equation}
        {\rm St_{drift}}=\frac{Z }{2 \eta}.
        \label{St_drift}
    \end{equation}
               
    On the other hand, when the dust particles' relative velocities exceed a specific threshold, they will fragment during collisions rather than perfectly merge. This sets another limit on the size of the particles, referred to as the fragmentation barrier. The fragmentation barrier in the turbulent velocity driven regime is given by 
    \begin{equation}
        {\rm St_{frag}}=\frac{v_{\rm frag}^2}{3\delta_{\rm t} c_{\rm s}^2},
        \label{St_frag}
    \end{equation}
    where \(v_{\rm frag}\) is the fragmentation threshold velocity and \(\delta_{\rm t}\) is the turbulent mixing parameter. In reality, the Stokes number of the largest particles can be approximated as $\rm St {=} min(St_{frag}, St_{drift})$. 

    It is worth noting that the turbulent mixing parameters are defined in three forms: $\delta_{\rm r}$, $\delta_{\rm t}$, and $\delta_{\rm z}$. For MRI-driven disk turbulence, which is approximately isotropic, angular momentum transport is roughly equivalent to mass diffusion \citep{johansen2005dust,zhu2012dust,Zhou_2022} ($\nu{\simeq}D_{\rm g}$), so that $\delta_{\rm r} {\simeq} \delta_{\rm t} {\simeq} \delta_{\rm z} {\simeq} \alpha$. In our default models, we set $\delta_{\rm r}{=}0, \delta_{\rm t} {\simeq} \delta_{\rm z} {\simeq} \alpha$, and  explore the influences of these parameters in Sect.~\ref{Sect.4.2}.

\subsection{Stellar X-ray photoevaporation}
\label{Sect.2.2}
    Intense X-ray radiation from the young, active central star heats the upper layers of the disk, inducing the hot gas and dust to escape as a photoevaporative wind \citep{alexander2014dispersal,ercolano2017dispersal}. In this work we focus on the stellar X-ray radiation, the influence of which is incorporated as the sink terms in Eq.~\ref{evo_g} and Eq.~\ref{evo_d}.
    
    We adopt the normalized \(\dot \Sigma_{\rm PE,g}\) profiles from \cite{Owen2012} for primordial disks and disks with inner holes. We also note that alternative prescriptions, such as that of \cite{Picogna2019}, yield different radial mass-loss distributions due to differences in their hydrodynamic setups and thermochemical treatments. More recently, \cite{sellek2024photoevaporation} find that the \cite{Owen2012}'s prescription remain in good agreement with their radiation-hydrodynamic simulations. This provides the motivation for our choice of the \cite{Owen2012} photoevaporation prescription in this work. Although these prescriptions differ in their detailed radial mass-loss profiles, a systematic exploration of how these differences may impact the disk dispersal and dust dynamics is beyond the scope of this study. 
    
    \cite{Owen2012} calculated the \(\dot \Sigma_{\rm PE,g}\) profiles for both primordial disks and disks with inner holes, based on their radiation-hydrodynamical model.
    \begin{equation}
        \dot{M}_{\rm PE} = 
            \begin{cases}
                1.65\times10^{-8}
                \left(\frac{M_\star}{M_\odot}\right)^{-0.068}
                \\
                \quad \quad \times\left(\frac{L_{\rm X}}{L_{\rm X,\odot}}\right)^{1.14}\ M_\odot\,\mathrm{yr^{-1}} \ \ &\text{[primordial disk]},
                \\
                \\
                1.27\times10^{-8}
                \left(\frac{M_\star}{M_\odot}\right)^{-0.148} &\text{[disk with inner hole,}
                \\
                \quad \quad \times\left(\frac{L_{\rm X}}{L_{\rm X,\odot}}\right)^{1.14}
                \ M_\odot\,\mathrm{yr^{-1}} &\,\,\text{at $r{>}r_{\rm hole}$]}.
            \end{cases}
        \label{mpe}
    \end{equation}
    
    The typical X-ray luminosity of a solar-mass T Tauri star can be expressed as \(L_{\rm X,\odot}{=}2.34\times10^{30}\ \rm erg\,s^{-1}\) \citep{preibisch2005}.
    Taking into account the correlation between X-ray luminosity of T Tauri stars and stellar mass (see Sect.4.2 in \citealt{preibisch2005}), the integrated gas mass-loss rate driven by the stellar X-ray photoevaporation can be written as
    \begin{equation}
        \dot{M}_{\rm PE} = 
            \begin{cases}
                1.65\times10^{-8}
                \left(\dfrac{M_{\star}}{M_\odot}\right)^{1.57}
                \ M_\odot\,\mathrm{yr^{-1}} \ \ &\text{[primordial disk]},
                \\
                \\
                1.27\times10^{-8}
                \left(\dfrac{M_{\star}}{M_\odot}\right)^{1.49}
                \ M_\odot\,\mathrm{yr^{-1}} \ \ &\text{[disk with inner hole,}
                \\ &\;\text{at $r{>}r_{\rm hole}$]}.
            \end{cases}
        \label{mpe_}
    \end{equation}

    For the primordial disks, the photoevaporative mass-loss rate $\dot{\Sigma}_{\rm PE,g}$ peaks at ${\sim}2{-}3$ au. After the cavity forms, we set the gas depletion rate at $r {<} r_{\rm hole}$ to a constant to ensure rapid gas removal there. Since the inner disk does not participate in planetesimal formation, this treatment does not alter the final mass of planetesimals. 
    We adopt a relatively high X-ray luminosity $L_{\rm X} {=} 6.15 \times 10^{30} \rm erg\,s^{-1} {=} 2.6\,L_{\rm X,\odot}$ in our default model, and assume that $L_{\rm X}$ remains constant during the short disk dispersal phase. Our simulations are designed to follow the final dispersal of the disk. Choosing a higher $L_{\rm X}$, corresponding to a more X-ray-active young star (see Fig. 3 in \citealt{preibisch2005}), allows the photoevaporation-driven clearing phase to complete within ${\sim}1$ Myr.

    The stellar photoevaporation also influences the dust dynamics. Following \cite{Facchini2016} and \cite{sellek2020}, we assume that  particles smaller than the entrained size, $a_{\rm ent}$, can be blown away by the wind, where
    \begin{equation}
        a_{\rm ent} = \frac{v_{\rm th}\dot{M}_{\rm PE} }{4\pi  \mathscr{F}\rho_{\rm \bullet} GM_{\star}},
        \label{a_ent}
    \end{equation}
    where \(v_{\rm th}{=}\sqrt{8/{\pi}}\,c_{\rm s}\) is the thermal velocity of the gas and \(\mathscr{F}{=}H_{\rm g}/{\sqrt{r^2+H_{\rm g}^2}}\) corresponds to the disk geometric factor. For particles smaller than the entrained size \(a_{\rm ent}\), we set the dust mass-loss rate 
    \begin{equation}
        \dot{\Sigma}^{i}_{{\rm PE,d}} = f_{\rm ent}\cdot Z^{i}\cdot \dot{\Sigma}_{\rm PE,g},
    \end{equation}
    where $\dot{\Sigma}^{i}_{\rm PE,d}$ is the dust mass-loss rate driven by the stellar X-ray photoevaporation within a specific dust mass bin $i$, $f_{\rm ent}$ is the entrained mass fraction, and $Z^{i}$ is the metallicity for that mass bin. Following \cite{sellek2020}, we set $f_{\rm ent}{=}1$ in our simulations.

\begin{table*}[t]
\caption{
List of simulation and model parameters used in Sect.~\ref{Sect.3}. 
}
\centering
\normalsize
\renewcommand{\arraystretch}{1.1} 
\begin{tabularx}{0.9\textwidth}{l *{7}{>{\centering\arraybackslash}X}}
    \toprule
    Model&$M_{\rm disk}\ [M_{\odot}]$&$R_{\rm c}$\ [au]&$\alpha$
    &$\delta_{\rm t}$&$Z_{\rm 0}$&$v_{\rm frag}\ [\rm m\,s^{-1}]$
    &$L_{\rm X}\ [L_{\rm X,\odot}]$\\
    \midrule
    \texttt{run\_fid}   &0.05 &60  &$10^{-4}$ &$10^{-4}$ &0.01 &5.0 &2.6\\
    \texttt{run\_metal} &0.05 &60  &$10^{-4}$ &$10^{-4}$ &0.02 &5.0 &2.6\\
    \texttt{run\_lumi}  &0.05 &60  &$10^{-4}$ &$10^{-4}$ &0.01 &5.0 &1.0\\
    \texttt{run\_alpha} &0.05 &60  &$10^{-3}$ &$10^{-4}$ &0.01 &5.0 &2.6\\
    \texttt{run\_dsize1} &0.10 &120 &$10^{-4}$ &$10^{-4}$ &0.01 &5.0 &2.6\\
    \texttt{run\_dsize2} &0.05 &120 &$10^{-4}$ &$10^{-4}$ &0.01 &5.0 &2.6\\
    \texttt{run\_frag}  &0.05 &60  &$10^{-4}$ &$10^{-4}$ &0.01 &1.0 &2.6\\
    \bottomrule
\end{tabularx}
\label{tab1}
\end{table*}

\begin{table*}[ht]
\caption{
Resultant planetesimal masses and dust-to-planetesimal conversion efficiencies.
}
\centering
\normalsize
\renewcommand{\arraystretch}{1.1} 
\begin{tabularx}{0.9\textwidth}{l *{6}{>{\centering\arraybackslash}X}}
        \toprule \multirow{2}{*}{Model} &
        \multicolumn{5}{c}{$M_{\rm plt}\ [M_{\oplus}]$} &
        \multirow{2}{*}{\makecell{Conversion\\ Efficiency\ [\%]}}\\
        \cmidrule(lr){2-6}
        & {1--3\ au} & {3--10\ au} & {10--30\ au} & {30--100\ au} & {Total} & \\
        \midrule
        \texttt{run\_fid}    &0    &14.0  &16.5  &0.9   &31.4   &20.4  \\
        \texttt{run\_metal}  &0    &26.2  &20.7  &1.3   &48.2   &15.3  \\
        \texttt{run\_lumi}   &0    &0     &0.8   &0.2   &1.0    &0.6   \\
        \texttt{run\_alpha}  &0    &5.9   &8.0   &0.7   &14.6   &9.5   \\
        \texttt{run\_dsize1} &0    &23.7  &73.8  &23.5  &125.0  &39.7  \\
        \texttt{run\_dsize2} &0    &9.7   &23.8  &28.9  &62.4   &40.5  \\
        \texttt{run\_frag}   &0    &0     &0     &0     &0      &0     \\
        \bottomrule
\end{tabularx}
\label{tab2}
\end{table*}

\subsection{Streaming instability criterion and planetesimal formation}
\label{Sect.2.3}
    Several literature studies have investigated the critical condition for triggering the SI. The classical criterion for the onset of the SI is proposed by \cite{youdin2005streaming} through a linear stability analysis. They found that the SI criterion can be fulfilled when the midplane volume density ratio of dust-to-gas approaches unity. Their result laid the foundation for subsequent theoretical and numerical studies. \cite{carrera2015form} explored this issue by performing $2$D numerical simulations and obtained the metallicity threshold $Z_{\rm crit}$ as a function of particles' Stokes number. Later \cite{yang2017concentrating} reduced this clumping threshold with an improved numerical method. \cite{li2021thresholds} further quantified how $\rm Z_{\rm crit}(St)$ vary with simulation setups. They found that sub-solar metallicity can already induce the dust clumping at $\rm St {\sim}0.1{-}1$, lower than previous results. However, their numerically study is only $2$D and without taking into account of disk turbulence. More recently, \cite{Lim_2024} advanced the understanding of the SI condition by performing $3$D stratified shearing box simulations that include both particles' self-gravity and externally forced turbulence. Their work provides a more accurate threshold for evaluating the planetesimal formation in global disk models. 

    For the purpose of generality, here we list three SI criteria, ranging from the early linear analysis of \cite{youdin2005streaming} to the most updated numerical simulations of \cite{Lim_2024}. The influence of adopting different SI criteria on our study will be discussed in Sect.~\ref{Sect.3.1}.
    
    To the zero order, the analytical SI criterion can be expressed as the midplane volume density ratio of dust-to-gas, where 
    \begin{equation}
        \epsilon_{\rm crit}\equiv \rho_{\rm d}/\rho_{\rm g} {=}1.
        \label{SI_clas}
    \end{equation}

    By explicitly quantifying how turbulence affects the SI-induced clumping, the new derived criteria from \cite{Lim_2024} bridge the gap between idealized SI theory and turbulent disk environments. The \(\epsilon_{\rm crit}\) criterion provided by
    \cite{Lim_2024} reads
    \begin{equation}
        \log_{10}\epsilon_{\rm crit} = 0.42(\log_{10}{\rm St})^2 + 0.72(\log_{10}{\rm St}) + 0.37.
        \label{SI_Lim24b_eps}
    \end{equation}
    Meanwhile, \citet{Lim_2024} also provided the $Z_{\rm crit}$ criterion such that
    \begin{equation}
        \begin{split}
            \log_{10}Z_{\rm crit} = &\ 0.15 \,(\log_{10}{\delta_{\rm t}})^2  -0.24 (\log_{10}{\rm St}\log_{10}{\delta_{\rm t}}) \\
            & -1.48(\log_{10}{\rm St}) + 1.18 (\log_{10}{\delta_{\rm t}}).
        \end{split}
        \label{SI_Lim24b_Z}
    \end{equation}
    The above two criteria hold for particles with  $\ 0.01 {\leq} {\rm St} {\leq} 0.1$ and disk turbulent level of $10^{-4}{\leq}\delta_{\rm t} {\leq}10^{-3}$.
    We also use YG05, L24E, and L24Z to represent Eqs.~\ref{SI_clas},~\ref{SI_Lim24b_eps}, and~\ref{SI_Lim24b_Z} hereafter.
   
    It is worth noting that L24Z (Eq.~\ref{SI_Lim24b_Z}) is derived at $\Pi{=}0.05$. Physically, a more decisive factor that sets the SI clumping should  actual be $Z/\Pi$ in the above equation \citep{sekiya2018two}.
    In our simulations that include the stellar X-ray photoevaporation, however, the $\Pi$ value varies significantly across the disk and crosses zero at the outer cavity edge. This variability makes it impractical to apply a correction to the criterion directly. We therefore adopt the $\epsilon_{\rm crit}$ criterion of L24E (Eq.~\ref{SI_Lim24b_eps}), which shares the same physical basis as the $Z_{\rm crit}$ criterion of L24Z (Eq.~\ref{SI_Lim24b_Z}), and subsequently calculate the midplane density ratio in our 1D code to incorporate the effects of $\Pi$. A key advantage of L24E (Eq.~\ref{SI_Lim24b_eps}) is its ability to capture turbulence variations from the SI (specifically through $\Pi$) via the self-consistent modification of the dust scale height. The specific method is detailed below. 

    In absence of the SI-induced dust concentration, the dust scale height adopted from \cite{dubrulle1995dust} can be written as 
    \begin{equation}
        H_{\rm d} = H_{\rm g}
        \sqrt{\frac{\delta_{\rm z}}{\delta_{\rm z}+{\rm St}}},
    \end{equation}
    where \(\delta_{\rm z}\) is the vertical turbulent mixing parameter. 
    SI can act as an additional turbulent source to affect the vertical dust distribution. Taking this into account, \cite{carrera2025positive} and \cite{eriksson2025} derived the effective dust scale height as 
    \begin{equation}
        H_{\rm d} = H_{\rm g}{\sqrt{\frac{1+{\rm St}}{1+{\rm St}+\epsilon}}} \sqrt{\left(\frac{\Pi}{5}\right)^2 + \frac{\delta_{\rm z}}{\delta_{\rm z}+{\rm St}}}.
        \label{adjs H_d}
    \end{equation}

    Our simulation is $1$D and only calculates the vertical integrated gas and dust quantities. In order to derive the volume density, the vertical distribution of the dust needs to be intrinsically assumed.    
    Despite that the dust vertical distribution might not be Gaussian-like owing to the SI effect \citep{Lim_2024}, we still assume that the dust midplane density approximately follows $\rho_{\rm d} {\propto} \Sigma_{\rm d}/ H_{\rm d}$, and therefore,     
    \begin{equation}
        \epsilon \equiv 
        \frac{\rho_{\rm d}}{\rho_{\rm g}} = Z \frac{H_{\rm g}}{H_{\rm d}}.
    \end{equation}
    Let \(\xi \equiv \frac{Z}{\sqrt{{1+{\rm St}}} \sqrt{\left(\frac{\Pi}{5}\right)^2+\frac{\delta_{\rm z}}{\delta_{\rm z}+{\rm St}}}}\),  we can obtain 
    \begin{equation}
        \epsilon = \frac{\xi^2+\sqrt{\xi^4+4\xi^2(1+{\rm St})}}{2}.
        \label{eps_t}
    \end{equation}

    We note that the aforementioned SI criteria are all derived from simulations based on single-size particles (see \citealt{bai2010dynamics,zhu2021streaming} for multi-species). In this study we consider the drift and coagulation of dust particles with a wide size range. We thus use the mass-weighted mean Stokes number to represent the bulk dust population, the formula of which can be written as
    \begin{equation}
        {\rm St_{mean}} = \frac{\sum_i {\rm St}^i\cdot\Sigma_{\rm d}^i}{\sum_i \Sigma_{\rm d}^i},
    \end{equation}
    where \({\rm St}^i\) and \(\Sigma_{\rm d}^i\) are the corresponding particles' Stokes number and mass density in the mass bin \(i\). The Stokes number of the dust population can therefore be specified at any given $t$ and $r$. We adopt ${\rm St_{mean}}$ for calculating $\epsilon_{\rm crit}$ and $H_{\rm d}$.

    Once $\epsilon{\geq}\epsilon_{\rm crit}$ is satisfied,  we transform the dust into planetesimals via
    \begin{equation}
        \dot{\Sigma}_{\rm plt}=- \zeta \frac{\Sigma_{\rm d}}{t_{\rm set}}=-\zeta\,\Sigma_{\rm d}\frac{\Omega_{\rm K}}{1+{\rm St}^{-1}},
        \label{plt form}
    \end{equation}
    where $t_{\rm set}{=}[\Omega_{\rm K}/(1{+}{\rm St}^{-1})]^{-1}$ is the settling time, which corresponds to the timescale on the formation of the SI filaments \citep{yang2017concentrating}.
    We adopt this formula of $t_{\rm set}$ to eliminate interference of particles with extremely large Stokes number, such as the particles within the inner cavity (see Fig.~\ref{fig:fiducial}b, Sect.~\ref{Sect.3.1}). 

    The planetesimal conversion efficiency, $\zeta$, represents the fraction of dust converted into planetesimals per settling time. Numerically, if $\zeta$ is too large, the entire dust mass within a radial bin would be depleted within a single timestep, leading to numerical instability. Physically, once the SI condition is met, dust trapped in a pressure bump is expected to rapidly transform into planetesimals within dozens of settling times ($t_{\rm set}$). Given that $t_{\rm set}$ is significantly shorter than the timescale for the movement of the photoevaporative cavity (${\sim}10^3{-}10^5$ yr),
    planetesimal formation should be efficient regardless of the adoption of $\zeta$ values across a broad parameter range.
    Based on the simulations of \cite{simon2016mass} and following \cite{schoonenberg2018lagrangian}, we set $\zeta{=}0.1$. 
    For completeness, we test several planetesimal formation prescriptions. Our fiducial model adopts the rate modified from \cite{schoonenberg2018lagrangian}. We further examine the formulations by \citet{drkazkowska2016}, which consider particles with $\rm St{>}0.01$ and an efficiency parameter $\zeta{=}0.01$, as well as the smooth transition function introduced by \citet{miller2021formation} to determine the local planetesimal formation fraction. The resulting differences in the total planetesimal mass are minor, suggesting that our results are robust against the specific choice of the planetesimal formation formula.
    
    We also note that although the SI criterion of L24E (Eq.~\ref{SI_Lim24b_eps}) is derived for particles with Stokes numbers in the range of ${\sim} 0.01{-}0.1$, in our study the dominated dust population involved in planetesimal formation naturally falls within a similar Stokes number range constrained by the drift and fragmentation barriers. The particles' mean Stokes number ($\rm St_{mean}$) reaches only approximately a few tenths even during the late stages of disk evolution (see Fig.~\ref{fig:fiducial}b, Sect.~\ref{Sect.3.1}). Practically, we extrapolate the thresholds beyond their original Stokes number range. We also test this approach and find that the difference is minor when restricting the Stokes number to ${\sim} 0.01{-}0.1$ compared to using the full size domain. 

\subsection{Simulation setup}
\label{Sect.2.4}
    We adopt the \texttt{DustPy (v1.0.5)} code \citep{Stammler_2022} to study the radial evolution of the gas and the dust in protoplanetary disk, with the effects of viscous gas evolution, stellar X-ray photoevaporation, dust coagulation, fragmentation and radial drift. The radial grid extends from $1{-}300\ \rm au$ with $256$ logarithmically spaced grid cells. 
    We additionally implement the local refinement of the grids in the planetesimal formation region driven by stellar X-ray photoevaporation, spanning \(\rm 5{-}20\ au\) in the typical simulation setups. 
    We verify that the final mass of planetesimals converges with this treatment and a higher radial grid resolution. 
    We adopt 120 bins for dust mass, logarithmically from $10^{-12}$ to $10^{5}\ \rm g$. The inner boundary condition for the dust and gas densities is set to be a constant gradient, whereas the outer boundary condition is set to be a floor value.

    There are six key parameters in our model: the initial metallicity $Z_0$, the stellar X-ray luminosity $L_{\rm X}$, the gas viscous parameter $\alpha$, the initial gas disk mass $M_{\rm disk}$, the disk characteristic disk radius $R_{\rm c}$, and the dust fragmentation threshold velocity $v_{\rm frag}$. The default parameter values are listed in Table~\ref{tab1}.

\section{Results}
\label{Sect.3}
In this section, we describe the results of the fiducial run in Sect.~\ref{Sect.3.1}, and explore different model parameters from Sect.~\ref{Sect.3.2} to Sect.~\ref{Sect.3.6}. Resultant planetesimal masses and dust-to-planetesimal conversion efficiencies are listed in Table~\ref{tab2}, and the radial mass distributions of planetesimals are presented in Fig.~\ref{fig:Mplt}.

\subsection{Fiducial model}
\label{Sect.3.1}
For the fiducial run (\texttt{run\_fid}), we choose initial disk mass \(M_{\rm disk}{=}0.05\ M_\odot\), characteristic disk radius \(R_{\rm c}{=}60\ \rm au\),  viscous parameter \(\alpha{=}10^{-4}\), initial metallicty \(Z_{0}{=}0.01\), fragmentation threshold velocity \(v_{\rm frag}{=}\rm 5\ m\,s^{-1}\), stellar X-ray luminosity \(L_{\rm X}{=}2.6\ L_{\rm X,\odot}\). The SI criterion is adopted from L24E (Eq.~\ref{SI_Lim24b_eps}).

\begin{figure}[ht]
    \centering
    \includegraphics[width=\linewidth]{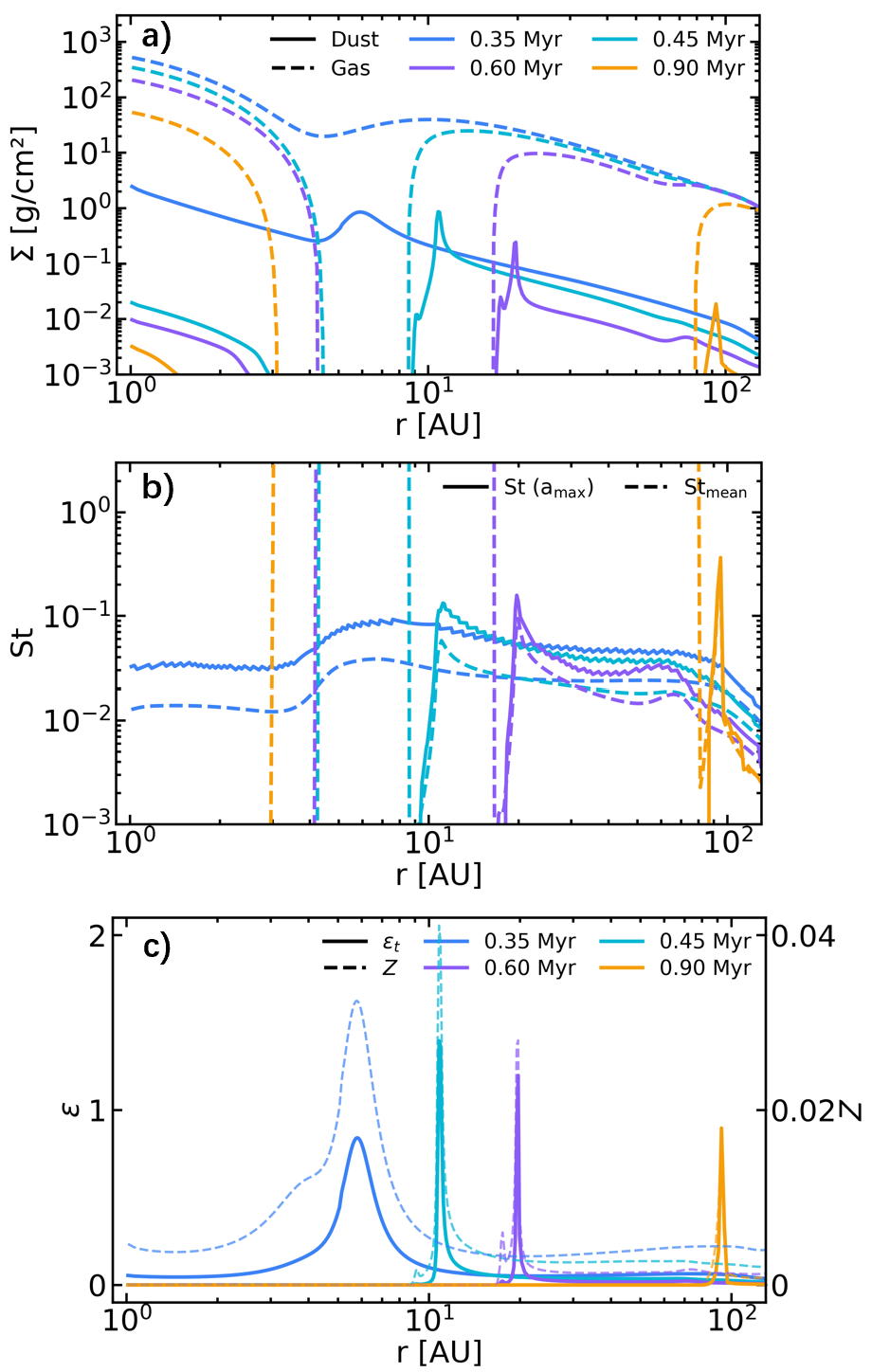}
    \caption{
    The time and radial distance evolution of different quantities in the fiducial model. 
    Panel a): Gas (dashed lines) and dust (solid lines) surface density. 
    Panel b): Stokes number of the maximum-size dust particles (solid lines) and the mean Stokes number of the dust particles (dashed lines) at each radial bin. Maximum-size refers to the size of the particle whose surface density is at the peak of the mass bins hereafter.
    Panel c): Midplane density ratio \(\epsilon\) (solid lines) and metallicity \(Z\) (dashed lines).
    }
    \label{fig:fiducial}
\end{figure}

Fig.~\ref{fig:fiducial}a shows the dust and gas disk evolution at different snapshots in \texttt{run\_fid}. At \(\rm t{=}0.35\ Myr\), an inner cavity opens at around $\rm 3 \ au$. The outer edge of this inner cavity expands outward rapidly, reaching $\rm {\sim} 10\ au$ at $\rm t{=}0.45\ Myr$, and $\rm {\sim} 100\ au$ at $\rm t{=}0.90\ Myr$, respectively. Due to the dust radial drift and gas disk inside-out clearing, the dust gets accumulated and forms a ring-like structure at the edge of the inner disk cavity.

\begin{figure*}[ht]
    \centering
    \includegraphics[width=\textwidth]{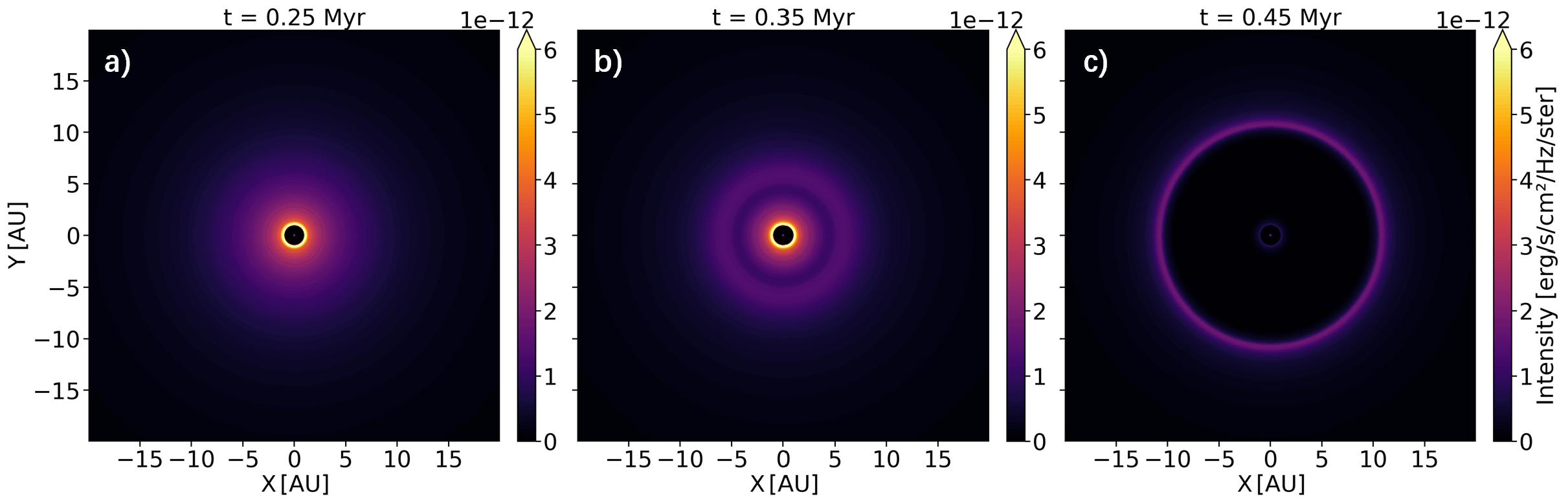}
    \caption{
    Synthetic \(\rm 880\ \mu m\) continuum images from the fiducial model.  
    Panel a): \(\rm t{=}0.25\ Myr\), pre-cavity stage; 
    Panel b): \(\rm t{=}0.35\ Myr\), onset of photoevaporative cavity opening; 
    Panel c): \(\rm t{=}0.45\ Myr\), post-cavity stage. 
    }
    \label{fig:880}
\end{figure*}

This can also be clearly seen in Fig.~\ref{fig:880}, which presents the synthetic dust continuum images at the \(\rm 880\ \mu m\) wavelength at three different epochs. At the early phase of $\rm t{=}0.25\ Myr$, the disk is driven by viscous evolution, and the dust is distributed relatively smoothly throughout the disk (see Fig.~\ref{fig:880}a). Fig.~\ref{fig:880}b shows the onset of photoevaporation-driven clearing, whereas in Fig.~\ref{fig:880}c, the inner cavity expands further out, with a strong dust emission at its outer edge. At the cavity's edge, rapid gas clearing yields a local pressure bump, providing a natural site for dust concentration. The images are generated with \texttt{RADMC-3D} based on the one-dimensional \texttt{DustPy} outputs mapped onto two-dimensional axisymmetric grids . We adopt 38 logarithmically spaced dust-size bins between $5\ \mu{\rm m}$ and $17\ {\rm cm}$, and assume that the disk is passively heated by stellar irradiation. Dust temperatures are computed using Monte Carlo radiative transfer, and synthetic images are convolved to $880\ \rm \mu m$, using isotropic scatterings and the DSHARP opacity model \citep{birnstiel2018disk}.

Fig.~\ref{fig:fiducial}b demonstrates the Stokes number of the maximum-size dust particles (solid lines) and the mean Stokes number of the dust population (dashed lines). Overall, the $\rm St\ (a_{max})$ and $\rm St_{mean}$ values are higher near the outer edge of cavity than in the residual disk regions. This is because dust drifts slower near the pressure bump, enhancing the local coagulation. We also obtain that \(\rm St_{mean}\) near the cavity's edge is \({\sim} 0.05\) at \(\rm t{=}0.45\ Myr\), \({\sim} 0.1\) at \(\rm t{=}0.60\ Myr\), and \({\sim} 0.2\) at \(\rm t{=}0.90\ Myr\). Thus, \(\rm St_{mean}\) remains largely below 0.1 within $30$ au, which is the most relevant for planetesimal formation. This size is within the range where L24E (Eq.~\ref{SI_Lim24b_eps}) is valid. 
We also note that the sharp increase in $\rm St_{mean}$ just inside the cavity is caused by the floor values of the surface densities set in \texttt{DustPy}. These anomalous Stokes number values, as discussed in Sect.~\ref{Sect.2.3}, can lead to the artificial formation of planetesimals in dust-depleted regions. Our adopted formula for $t_{\rm set}$ (Eq.~\ref{plt form}) is designed to prevent this unphysical outcome.

Fig.~\ref{fig:fiducial}c illustrates the ratios of dust-to-gas volume density $\epsilon$ and surface density $Z$. Due to dust trapping at the pressure bump, the disk metallicity $Z$ gets largely elevated and exceeds $2{-}4\%$, whereas $\epsilon$ reaches the order of unity at the edge of the inner cavity as it sweeps outward.

We adopt the SI criterion from L24E (Eq.~\ref{SI_Lim24b_eps}). In the end, planetesimals with a total mass of $31.4\ M_\oplus$ are generated through this photoevaporation gas clearing process. The dust-to-planetesimal conversion efficiency is 20.4\%, which indicates that 20.4\% of the initial dust mass has been converted into planetesimals. The influences of alternative criteria of YG05 (Eq.~\ref{SI_clas}) and L24Z (Eq.~\ref{SI_Lim24b_Z}) on planetesimal formation are also tested. The total masses of resultant planetesimals show minor differences among all these criteria (${\lesssim}10\%$). Thus, the \(\epsilon_{\rm crit}\) criterion of L24E is used as the default in all following simulations.

\begin{figure*}[ht]
    \centering
    \includegraphics[width=\linewidth]{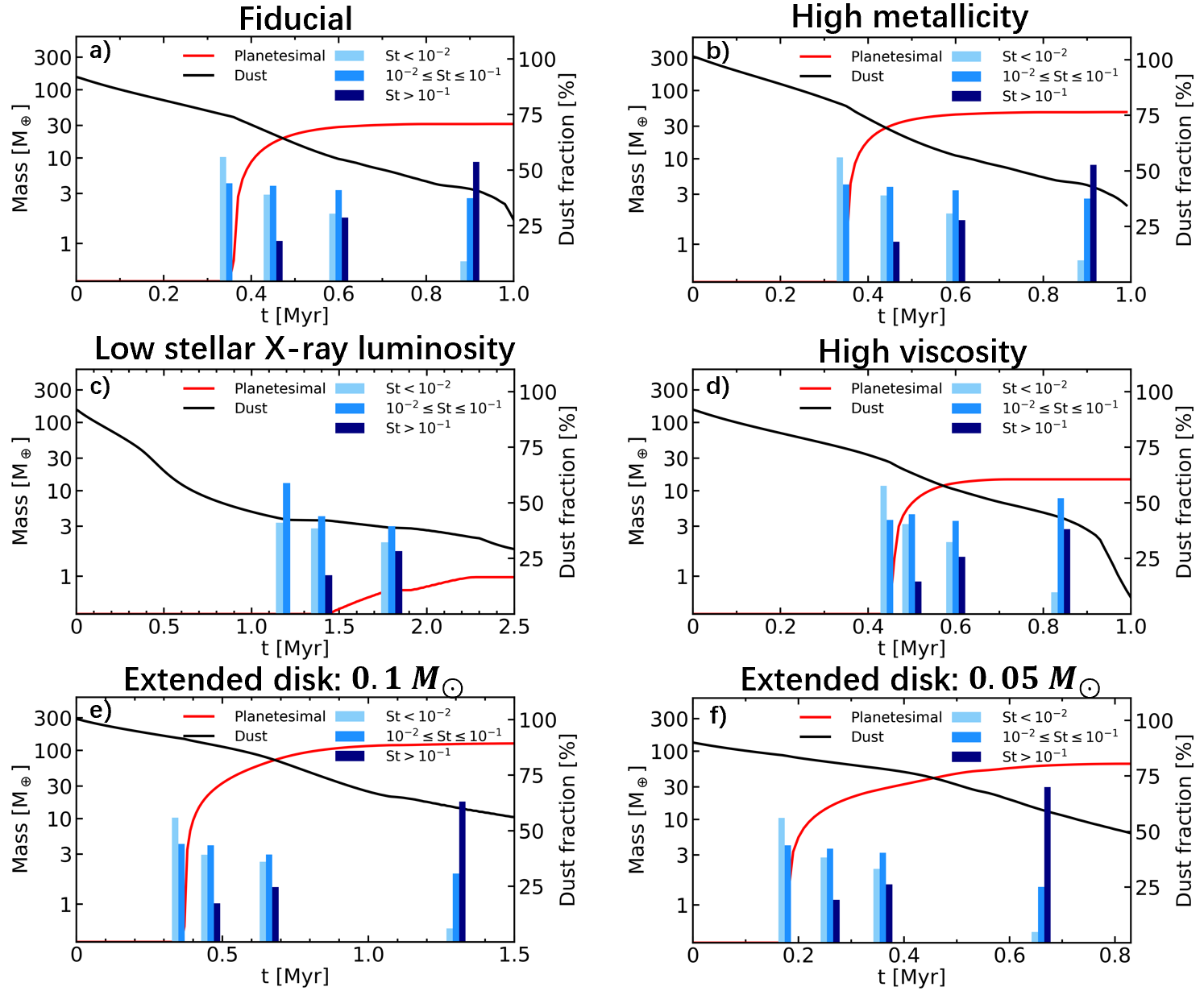}
    \caption{
    The evolution of dust mass (black lines), planetesimal mass (red lines), and dust mass fraction within the pressure bump in three Stokes number bins (light‑blue: \(\rm{St}{<}10^{-2}\); blue: \(10^{-2}{\leq}\rm{St}{\leq}10^{-1}\); deep‑blue: \(\rm{St}{>}10^{-1}\)), across four evolutionary stages: (i) pre‑cavity, (ii) cavity at 10 au, (iii) cavity at 20 au, and (iv) cavity at 100 au. 
    These panels show how initial conditions affect dust retention and planetesimal formation: 
    Panel a) Fiducial model (\(Z=0.01\), \(L_{\rm X}=2.6\,L_{\rm X,\odot}\), \(M_{\rm disk}=0.05\,M_\odot\), \(R_{\rm c}=60\,\rm au\), \(\alpha=10^{-4}\)) serves as the baseline for comparison, showing balanced dust evolution and planetesimal production across the disk. 
    Panel b) Higher metallicity (\(Z=0.02\)) facilitates the retention of dust mass, thereby promoting planetesimal formation. 
    Panel c) Decreased X‑ray luminosity (from \(2.6\ L_{\rm X,\odot}\) to \(1.0\ L_{\rm X,\odot}\)) diminishes dust mass, and consequently impedes planetesimal formation. 
    Panel d) Higher viscosity parameters (\(\alpha{=}10^{-3}\), \(\delta_{\rm t}{=}10^{-4}\)) induce gas replenishment from the outer disk, delay cavity opening, reduce overall dust reservoir, and thus hinder planetesimal formation. 
    Panel e) A more massive and extended disk (\(M_{\rm disk}{=}0.1\ M_\odot\), \(R_{\rm c}{=}\rm 120\ au\)) prolongs dust retention, thereby extending the duration of planetesimal formation. 
    Panel f) A less massive but extended disk (\(M_{\rm disk}=0.05\,M_\odot\), \(R_{\rm c}=120\,\rm au\)) lowers the inner-disk dust density, causing photoevaporation to open a cavity earlier. This makes more dust retained at the onset of photoevaportative cavity, which also provides more building blocks for planetesimal formation. 
    }
    \label{fig:M-t}
\end{figure*}

Fig.~\ref{fig:M-t}a shows the evolution of the total dust mass (black line) and the planetesimal mass (red line) in \texttt{run\_fid}. The blue bars represent the mass distribution of dust across different size bins within the pressure bump. The rapid rise of the planetesimal mass (red line) indicates that trapped dust particles are efficiently converted into planetesimals. This high formation efficiency results from the midplane volume density ratio $\epsilon$ being sufficiently high to trigger the SI at the cavity edge. Furthermore, after cavity opening, the dust population within the pressure bump is predominantly composed of particles with Stokes numbers of $0.01{-}0.1$. The contribution of the dust particles within the pressure bump to the final planetesimal mass depends solely on the mass fraction of the trapped dust particles.

To conclude, \texttt{run\_fid} shows that stellar X-ray photoevaporation opens a cavity at a few au orbital distance. As the inner cavity sweeps inside-out, the dust particles accumulating  at the cavity's edge can trigger the SI and form planetesimals efficiently. In total $30.5 \ M_{\oplus}$ of planetesimals are generated at $r{=}5{-}30$\ au within $1$ Myr after the onset of stellar X-ray photoevaporation.

\begin{figure*}[h]
    \centering
    \includegraphics[width=\textwidth]{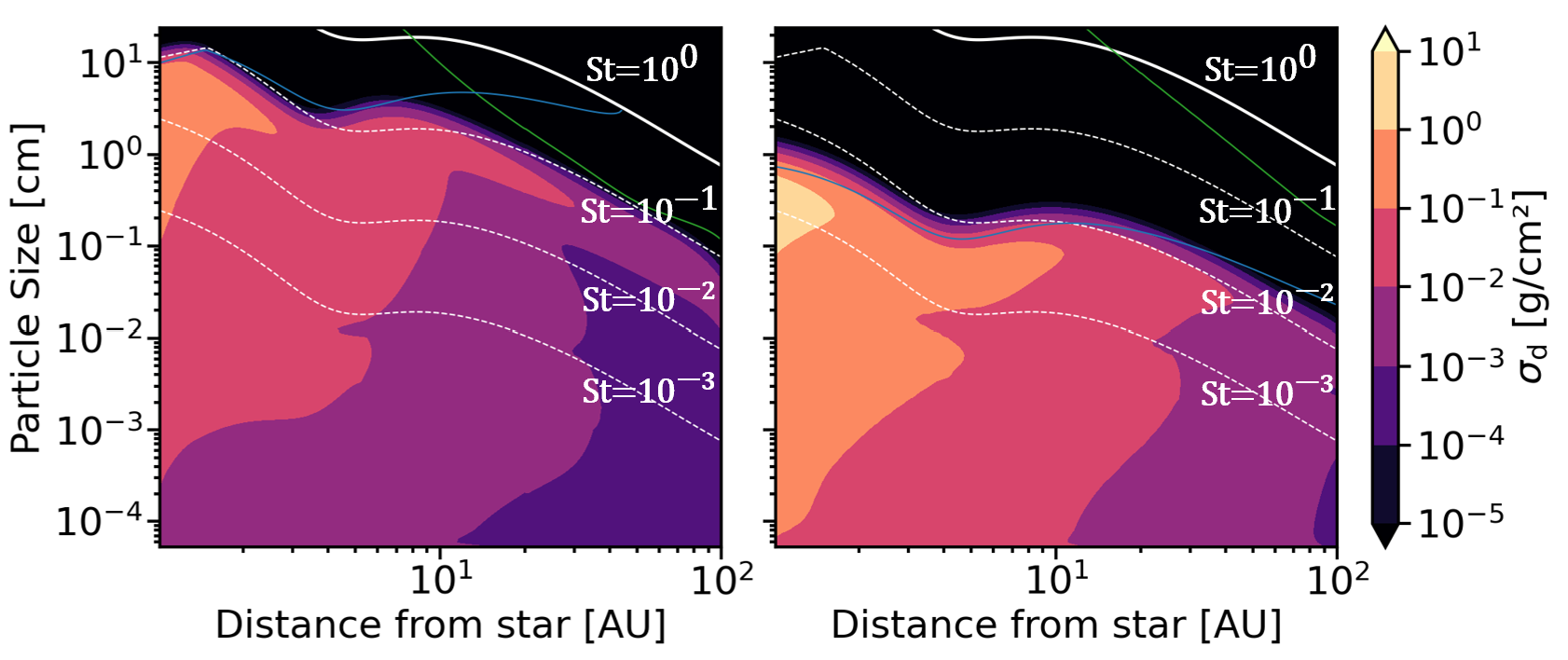}
    \caption{The distribution of dust as a function of radial distance and particle size at $t{=}0.30\ \rm Myr$ for \texttt{run\_fid} (left) and \texttt{run\_frag} (right). The color bar represents the surface density of dust per logarithmic mass interval. The green line is the drift limit (Eq.~\ref{St_drift}), whereas blue line is the fragmentation limit (Eq.~\ref{St_frag}). The four white lines correspond to $\rm St{=}[10^{-3}, 10^{-2}, 10^{-1}, 10^{0}]$}
    \label{fig:v_frag}
\end{figure*}

\subsection{Metallicity}
\label{Sect.3.2}
    We investigate the influence of disk metallicity on planetesimal formation, by increasing $Z_0$ from 0.01 in \texttt{run\_fid} to 0.02 in \texttt{run\_metal}. The other parameters are kept identical as \texttt{run\_fid}. 

    Fig.~\ref{fig:M-t}b shows total dust mass and planetesimal mass in \texttt{run\_metal}. The initial dust mass is \({\sim}300\ M_\oplus\), whereas \(50\ M_\oplus\) remains when the photoevaporation opens a cavity at \(\rm t{=}0.37\ Myr\). 
    For comparison, in \texttt{run\_fid}, the initial dust mass is \({\sim}150\ M_\oplus\), whereas \(37\ M_\oplus\) remains at $ t{=}0.37 $ Myr ( see Fig.~\ref{fig:M-t}a).
    
    We find that \texttt{run\_fid} yields a final planetesimal mass of \(31.4\ M_\oplus\), 
    whereas in \texttt{run\_metal} the final planetesimal mass reaches \(48.2\ M_\oplus\). 
    Clearly, doubling the initial metallicity indeed results in an increase of the final planetesimal mass. But the final planetesimal mass in \texttt{run\_metal} is not twice as high as in \texttt{run\_fid}. This is because the dust in the \texttt{run\_metal} experiences more rapid growth (see Eq.~\ref{tau_grow}) compared to that in \texttt{run\_fid}. In the early phase, the grown particles also get accreted by the central star at a faster rate. As a consequence, in both cases the masses of retained dust at the onset of photoevaporative cavity are overall similar. 

\subsection{Stellar X-ray luminosity}
\label{Sect.3.3}
    Meanwhile, we also test the effect of the stellar X-ray luminosity such that $L_{\rm X}$ is set to be \(1.0\ L_{\rm X,\odot}\) in \texttt{run\_lumi} rather than \(2.6\ L_{\rm X,\odot}\) in \texttt{run\_fid}. The other parameters are the same as \texttt{run\_fid}. 
    
    Table~\ref{tab2} shows that \texttt{run\_fid} yields a final planetesimal mass of \(31.4 \ M_\oplus\) with an overall dust-to-planetesimal conversion efficiency of 20.4\%, whereas \texttt{run\_lumi} achieves a final planetesimal mass of only \(1\ M_\oplus\) with a low conversion efficiency of 0.6\%. As can be seen from Fig.~\ref{fig:M-t}c, only \(~3.8\ M_\oplus\) of dust is left when the photoevaporation opens a cavity at \(\rm t{=}1.2\ Myr\) in \texttt{run\_lumi}. 
    
    The low stellar X-ray luminosity model results in a significant drop in the planetesimal mass and conversion efficiency. The strength of $L_{\rm X}$ is crucial to determine the time for the photoevaporation-driven cavity formation. The disk around a star with a low stellar X-ray luminosity results in the cavity-opening at a later time. By this time, a substantial amount of dust has already been accreted to the star. As a consequence, only a limited amount of dust remains available for the planetesimal formation. Therefore, we suggest that a relatively high $L_{\rm X}$ is required to induce the generation of planetesimals at the gas clearing phase.

\subsection{Viscosity}
\label{Sect.3.4}
    Disk viscosity, paramterized by \(\alpha\), drives the global evolution of the protoplanetary disk. 
    We focus on isolating the role of viscosity $\alpha$ in global disk evolution and that of the turbulent parameter $\delta_{\rm t}$ in dust coagulation. In other words, we treat $\alpha$ and $\delta_{\rm t}$ differently. In \texttt{run\_alpha} we only vary \(\alpha\) from $10^{-4}$ to $10^{-3}$ whereas the turbulent mixing parameter \(\delta_{\rm t}\) is still kept the same. Thus, the particles' Stokes number set by the fragmentation limit is the same as in \texttt{run\_fid}.
    
    Fig.~\ref{fig:M-t}d reveals that at a high $\alpha$ of $10^{-3}$, owing to more efficient angular momentum transport, the gas from the outer disk region replenishes the inner disk region at a fast pace.
    The cavity is opened at a later time of $0.45$ Myr in \texttt{run\_alpha}. 
    As a result, a smaller amount of dust is left by this time and eventually lead to less planetesimal formation. Table~\ref{tab2} shows that the planetesimal mass decreases from \(31.4\ M_\oplus\) in \texttt{run\_fid} to \(14.6\ M_\oplus\) in \texttt{run\_alpha}. 

\subsection{Disk radius}
\label{Sect.3.5}
    The disk size is also a key factor that affects the gas/dust evolution and planetesimal formation.  In \texttt{run\_fid}, the characteristic disk radius \(R_{\rm c}\) is set to be \(\rm 60\ au\).
    Here we perform two additional sets of simulations to explore the influence of disk size. 
    In \texttt{run\_dsize1}, we keep $\Sigma_{\rm g0}$ at $1$ au the same as that of \texttt{run\_fid} and increase \(R_{\rm c}\) to \(\rm 120\ au\). In such a case the initial disk mass increases to \(0.1\ M_\odot\). On the other hand, in \texttt{run\_dsize2}, the total disk mass is still kept the same as \(0.05\ M_\odot\) whereas \(R_{\rm c}\) is increased to \(\rm 120\ au\). This configuration leads to an overall lower $\Sigma_{\rm g0}$ in the inner disk region. 

    Fig.~\ref{fig:M-t}e \& f  depict the evolution of dust masses and planetesimal masses in these two cases. The disk is more massive and a higher proportion of dust mass resides in the outer disk region in \texttt{run\_dsize1} compared to \texttt{run\_fid}. Since the gas density remains approximately the same in the inner disk region for \texttt{run\_fid} and \texttt{run\_dsize1},  the photoevaporative cavity forms at a similar time. In \texttt{run\_dsize1}, the residual dust mass reaches  \({\sim}144\ M_\oplus\) at this cavity-opening time (see Fig.~\ref{fig:M-t}e). This leads to a higher planetesimal mass and conversion efficiency in the end.

    On the other hand, \texttt{run\_dsize2} refers to a circumstance that the gas mass is more readily distributed at further out disk region. For \texttt{run\_dsize2}, the gas density in the inner disk region is lower, and the  photoevaporation opens a cavity at an earlier time of $0.18$ Myr compared to \texttt{run\_fid} (see Fig.~\ref{fig:M-t}f). This makes more dust retained at the onset of photoevaportative cavity, which also provides more building blocks for planetesimal formation. 

    As can be seen from Table~\ref{tab2}, the conversion efficiencies in above two models are quite similar, but both much higher than \texttt{run\_fid}. This again, indicates that regardless of the detailed treatment of $\Sigma_{\rm g}$ or $M_{\rm disk}$, a disk with a larger size promotes the dust retention and thus subsequent planetesimal formation.

\subsection{Dust fragmentation threshold velocity}
\label{Sect.3.6}
    In order to explore how the photoevaporation-driven planetesimal formation depends on the fragmentation-limited dust growth, we perform a simulation \texttt{run\_frag} with a lower fragmentation threshold velocity of \(v_{\rm frag} {=} \rm 1\ m\, s^{-1}\). The other parameters are the same as \texttt{run\_fid}. 
  
    Fig.~\ref{fig:v_frag} shows the distribution of dust as a function of radial distance and particle size at $t{=}0.30\rm \ Myr$ for \texttt{run\_fid} (left) and \texttt{run\_frag} (right). It is evident that the particles' Stokes number in \texttt{run\_fid} is more than one order of magnitude larger than that in \texttt{run\_frag}. Because the growth is limited by fragmentation, the largest particles reach a lower Stokes number when the fragmentation threshold is lower. Such small particles are more strongly coupled to the disk gas, and hence drift more slowly. Consequently, they are less efficient to concentrate at the cavity's edge, preventing the SI clumping and subsequent planetesimal formation. As can be seen from Table~\ref{tab2}, no planetesimal can form during the gas clearing phase in \texttt{run\_frag}.
    
    \cite{lim2025probing} and \cite{lim2025streaming} conducted $2$D and $3$D simulations of pure SI without external turbulence, respectively. The SI criteria presented in these two studies suggested lower thresholds for the onset of the SI. However, it is worth noting that their simulations indicate that $\epsilon$ needs to exceed 2 to trigger the SI for particles with $\rm St{\leq}10^{-3}$. Hence, even adopting a loose SI criterion from \cite{lim2025streaming} (their Eq. 11) without considering any turbulence, planetesimals are still unlikely to form at a relatively low fragmentation velocity. \cite{zhao2025planetesimal} also concluded that $v_{\rm frag}{=}2\ \rm m\,s^{-1}$ was the minimum value that allowed planetesimal formation, which is consistent with our results.

\begin{figure}[ht]
    \centering
    \includegraphics[width=\linewidth]{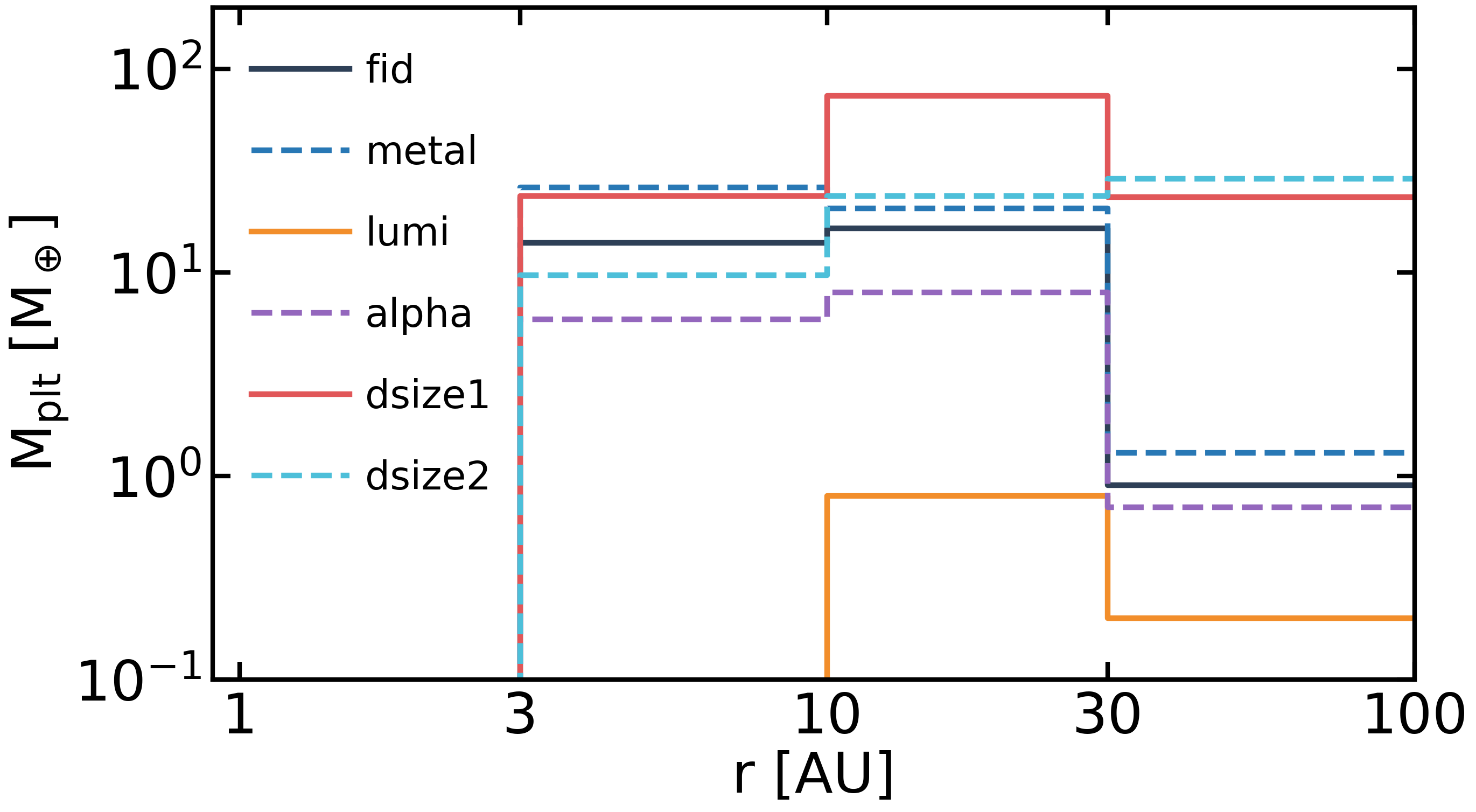}
    \caption{Comparison of the radial mass distributions of planetesimals at different disk regions. The curves show the results for models with varying parameters: \texttt{run\_fid} (black), \texttt{run\_metal} (blue), \texttt{run\_lumi} (orange), \texttt{run\_alpha} (purple), \texttt{run\_dsize1} (red), and \texttt{run\_dsize2} (cyan).}
    \label{fig:Mplt}
\end{figure}

    To facilitate a direct comparison among different models, Fig.~\ref{fig:Mplt} presents the radial mass distributions of planetesimals. To conclude, our results indicate that planetesimal formation is more efficient in larger disks with higher metallicities, lower viscosities, higher dust fragmentation threshold velocities, and/or around stars with higher X-ray luminosities.

\section{Discussions}
\label{Sect.4}
We compare our work with the literature studies in Sect.~\ref{Sect.4.1}. The influences of turbulent mixing parameters and pressure bumps produced by other mechanisms are presented in Sect.~\ref{Sect.4.2} and Sect.~\ref{Sect.4.3}. Finally, the caveats in our model are discussed in Sect.~\ref{Sect.4.4}.

\begin{table*}[t]
\caption{List of simulation setup, model parameters and resultant planetesimal masses in Sect. \ref{Sect.4.2}.}
\centering
\normalsize
\renewcommand{\arraystretch}{1.1} 
\begin{tabularx}{0.9\textwidth}{l *{6}{>{\centering\arraybackslash}X}}
    \toprule
    Model&$\alpha$&$\delta_{\rm r}$&$\delta_{\rm t}$
    &$\delta_{\rm z}$&$v_{\rm frag}\ [\rm m\,s^{-1}]$&$M_{\rm plt}\ [M_\oplus]$\\
    \midrule
    \texttt{run\_fid}    &$10^{-4}$ &0       &$10^{-4}$ &$10^{-4}$ &5   &31.4\\
    \texttt{run\_alpha}  &$10^{-3}$ &0       &$10^{-4}$ &$10^{-4}$ &5   &14.6\\
    \texttt{run\_rad}    &$10^{-4}$ &$10^{-4}$ &$10^{-4}$ &$10^{-4}$ &5   &30.9\\
    \texttt{run\_turb}   &$10^{-3}$ &0       &$10^{-3}$ &$10^{-4}$ &15  &0.7 \\
    \texttt{run\_vert}   &$10^{-4}$ &0       &$10^{-4}$ &$10^{-6}$ &5   &20.9\\
    \bottomrule
\end{tabularx}
\label{tab3}
\end{table*}

\subsection{Comparison with literature studies}
\label{Sect.4.1}
    We compare our results with those of \cite{carrera2017planetesimal} and \cite{ercolano2017x}, who also explored planetesimal formation via the SI driven by the stellar X-ray photoevaporation.

    First, our treatment of dust evolution differs from theirs. Whereas \cite{carrera2017planetesimal} used a custom 1D viscous disk code with simplified dust prescriptions from \cite{birnstiel2011dust}, and \cite{ercolano2017x} employed the two-population model of \cite{birnstiel2012simple} in their $1$D code, we utilize the open-source \texttt{DustPy} code, which incorporates full multi-size dust evolution, including coagulation, fragmentation, and radial drift. The other difference is the treatment of stellar X-ray photoevaporation. \cite{carrera2017planetesimal} adopted the model from \cite{gorti2015impact}, who considered FUV-driven disk dispersal. This process triggers strong radial pressure gradients and enhance the dust pile-ups in the early phase of the disk lifetime. \cite{ercolano2017x}, by contrast, used both X-ray and EUV-driven photoevaporation prescription derived from \cite{owen2010radiation,Owen2012}'s hydrodynamic simulations. Consequently, in their works, disk gas was removed more gradually, and photoevaporation primarily affected the inner a few au region of the disk. While also adopting the stellar X-ray photoevaporation prescription from \cite{Owen2012}, we additionally incorporate dust removal via photoevaporation, following the studies of \cite{Facchini2016} and \cite{sellek2020}.

    There are two improvements in our study: the dust scale height calculation and SI criterion. Whereas the previous studies relied on simplified estimates for the midplane density, we implement the correction from \cite{carrera2025positive} and \cite{eriksson2025} (our Eq.~\ref{adjs H_d}), which self-consistently accounts for local turbulence and particle-size-dependent settling. This allows us to track dust evolution more precisely and significantly improves the accuracy of midplane density predictions. 
    On the other hand, both \cite{carrera2017planetesimal} and \cite{ercolano2017x} adopted the classical threshold from the 2D unstratified simulations of \cite{carrera2015form}. In this study, we instead adopt the latest criterion from \cite{Lim_2024}, which is based on 3D shearing box simulations and defines a critical midplane density ratio \(\epsilon_{\rm crit}\). This criterion takes into account the external turbulence, reflecting a more realistic condition for dust-gas interaction in vertically stratified disks.

    Regarding the planetesimal formation, \cite{carrera2017planetesimal} reported up to \(60\ M_\oplus\) of planetesimals formed beyond 100 au region, and ${\sim} 16\ M_\oplus$ within 100 au. This high efficiency is a consequence of their adoption of strong FUV photoevaporation, which accelerates disk dispersal in the outer disk region and thus enhances dust concentration there. Conversely, \cite{ercolano2017x} found planetesimal formation is less likely for the X-ray and EUV photoevaporation. Their model produces only $3\ M_\oplus$ under optimal conditions, due to rapid dust drift before gap formation and absence of strong pressure traps. Our results strike a middle ground. We demonstrate that stellar X-ray photoevaporation can still trigger localized SI when vertical dust settling is explicitly modeled. Because X-ray–driven winds have higher temperatures and thus smaller gravitational radii, the photoevaporative cavity opens at a smaller physical radius, allowing a larger fraction of solids to remain inside ${\sim}100$ au and thereby promoting planetesimal formation within ${\sim}100$ au.
    In our fiducial model, the total planetesimal mass can attain \(30.5\ M_\oplus\) within the $30$ au disk region.
    By adopting a more realistic SI criterion and a sophisticated dust evolution treatment, our study indicates that stellar X-ray photoevaporation can also be robust for SI-triggered planetesimal formation.

\subsection{Turbulent mixing parameters}
\label{Sect.4.2} 
    Several turbulent-relevant parameters are used in this work: \(\delta_{\rm r}\), \(\delta_{\rm t}\) and \(\delta_{\rm z}\). Since disk turbulence can be non-isotropic, e.g., driven by the vertical shear instability (VSI) \citep{nelson2013linear} or non-ideal MHD \citep{yang2018diffusion}, these parameters are not necessarily equal. As shown in Table~\ref{tab3}, we conduct several simulations to investigate the influences of varying these parameters.

    We perform a simulation \texttt{run\_rad}, with considering the effect of turbulent radial diffusion on dust and other parameters kept identical as \texttt{run\_fid}. The dust diffusion timescale in \texttt{run\_rad} is given by
    \begin{equation}
        \begin{aligned}
            t_{\rm diff} = \frac{(\Delta r)^2}{D_{\rm d}}
            \simeq 1.5\times10^{3} f^2 \left(\frac{\delta_{\rm r}}{10^{-4}}\right)^{-1} 
            \left(\frac{P_{\rm r}}{1\ \rm yr}\right)\ \rm yr,
        \end{aligned}
    \end{equation}
    where $\Delta r {\sim} fH_{\rm g}$ is the pressure bump width at the onset of photoevaporation induced cavity, $f$ is an order-unity factor, and $P_{\rm r}$ is the Keplerian period at the radial distance $r$. 
    
    The photoevaporative cavity expansion timescale can be estimated as
    \begin{equation}
        \begin{aligned}
            t_{\rm cav} = \frac{\Delta r}{v_{\rm cav}} = \frac{\Delta r}{r}\ \frac{\Sigma_{\rm cav}}{\dot{\Sigma}_{\rm PE,g}} 
            \simeq 10^{3} f^2 \left(\frac{\alpha}{10^{-4}}\right)^{-1} 
            \left(\frac{P_{\rm r}}{1\ \rm yr}\right)\ \rm yr,
        \end{aligned}
    \end{equation}
    where $v_{\rm cav}{=}r \dot{\Sigma}_{\rm PE,g} / \Sigma_{\rm cav}$ is the cavity expansion velocity, we emphasize that $v_{\rm cav}$ does not correspond to the physical radial velocity of the gas; rather, it represents the expansion speed (i.e., the phase velocity) of the photoevaporative cavity edge. $\Sigma_{\rm cav} {\simeq} \dot{M}_{\rm acc} / (3 \pi \nu)$ denotes the critical gas surface density at which viscous accretion balances photoevaporative mass loss, and $\dot{M}_{\rm acc} {\simeq} 2 \pi r \Delta r \dot{\Sigma}_{\rm PE,g}$ corresponds to the critical gas accretion rate.
    
    We note that, in estimating $t_{\rm cav}$, we have assumed $\dot{M}_{\rm acc} {\simeq} 2 \pi r \Delta r \dot{\Sigma}_{\rm PE,g}$ for simplicity. In practice, however, the gas accretion rate after the onset of cavity opening is typically smaller than the photoevaporative mass-loss rate. This implies that our estimate of $t_{\rm cav}$ should be regarded as an upper limit. Therefore, we provide only an estimate that $t_{\rm cav} {\lesssim} t_{\rm diff}$. Although the exact reduction factor is difficult to quantify, this effect further reinforces our conclusion that radial diffusion cannot regulate the dust concentration at the cavity edge.
    
    The dust particles settle and are subsequently converted into planetesimals on a timescale of
     \begin{equation}
        t_{\rm plt} = \frac{1}{\zeta \rm St \Omega_K } \simeq 16 \  \left( \frac{\zeta}{0.1} \right)^{-1}  \left( \frac{\rm St}{0.1} \right)^{-1} \left( \frac{P_{\rm r}}{1 \rm \  yr} \right)
        \ \rm yr,
    \end{equation}
    where $\rm St$ is the Stokes number of the maximum-size dust particles $\rm St\ (a_{max})$, which follows $\rm St\ (a_{max}){\simeq}0.1 (r/\rm au)^{1/2}$ in the fiducial model.

    It can be noted that, for the particles with Stokes numbers of $0.01{-}0.1$, $t_{\rm plt}{\ll}t_{\rm cav}{<}t_{\rm diff}$ in \texttt{run\_rad}. Once the SI criterion is met, dust is converted into planetesimals on a timescale much shorter than both the outward expansion of the photoevaporative cavity edge and the radial dust diffusion timescale. In other words, planetesimal formation is completed before the cavity edge recedes and before diffusion can significantly redistribute the dust. Consequently, we find no substantial difference in the total planetesimal mass formed between between \texttt{run\_rad} and \texttt{run\_fid} (see Table~\ref{tab3}).

    We set $\delta_{\rm r}{=}0$, since radial diffusion does not regulate the peak dust-to-gas ratio in the low $\delta_{\rm r}$ and high $L_{\rm X}$ regime we focus on here. In this regime, photoevaporation drives the cavity edge outward on a timescale shorter than that required for radial dust diffusion to establish a local drift–diffusion equilibrium at the pressure maximum. More importantly, both $t_{\rm cav}$ and $t_{\rm diff}$ are much longer than $t_{\rm plt}$. Once the SI is triggered, dust is rapidly converted into planetesimals, which naturally limits the peak dust-to-gas ratio.
    
    Having explored the influence of the radial diffusion mechanism on planetesimal formation, we now turn our attention to another important parameter, \(\delta_{\rm t}\). Furthermore, in order to investigate the influence of \(\delta_{\rm t}\), we conduct a control simulation, \texttt{run\_turb}, based on  \texttt{run\_alpha}. As shown in Table~\ref{tab3}, this control run maintains all other parameters the same as \texttt{run\_alpha}, but adjusts \(\delta_{\rm t}\) to \(10^{-3}\) and \(v_{\rm frag}\) to \(\rm 15\ m\,s^{-1}\). This ensures that the dust reach a similar Stokes number at the fragmentation limit in two simulations.
    
    We can therefore isolate the effect of \(\delta_{\rm t}\). Through this comparison, we find that an increase in \(\delta_{\rm t}\) leads to a reduction of final planetesimal mass (from 14.6 \(M_\oplus\) to 0.7 \(M_\oplus\)). This phenomenon can be attributed to the fact that the relative velocity between particles increases with \(\delta_{\rm t}\), and dust grows much faster in this circumstance (see Eq.~\ref{tau_grow}). Consequently, the dust in the inner disk reaches the fragmentation limit at an earlier stage, leading to a less efficient retention of dust mass  before the onset of photoevaprative cavity.

\begin{figure*}[ht]
    \centering
    \includegraphics[width=\textwidth]{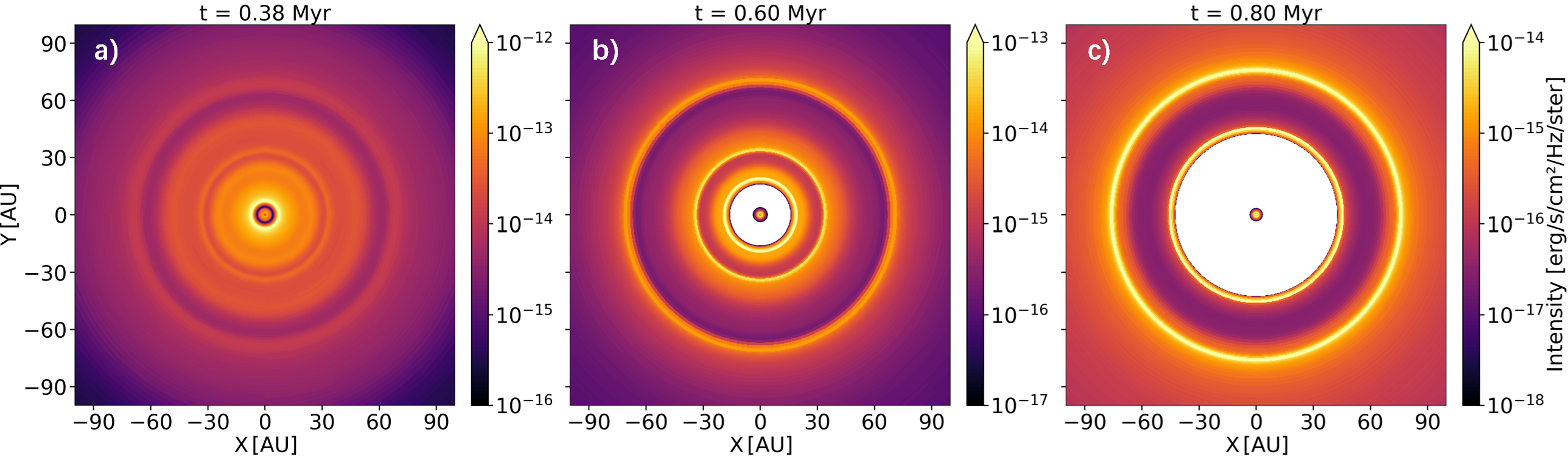}
    \caption{Synthetic dust continuum images at \(\rm 880\ \mu m\) for three evolutionary stages of the protoplanetary disk model with three local pressure bumps at 30 au and 60 au in Sect.~\ref{Sect.4.3}. Panel a): \(\rm t{=}0.38\ Myr\), pre-cavity stage; Panel b): \(\rm t{=}0.60\ Myr\), post-cavity stage when the photoevaporative cavity moves to 18 au; Panel c): \(\rm t{=}0.80\ Myr\), when the photoevaporative cavity merges with the inner bump at 30 au. The blanks within the inner disks in Panel b) and c) correspond to the photoevaporative cavities.}
    \label{fig:bump}
\end{figure*}

    Similar to \(\delta_{\rm t}\), \(\delta_{\rm z}\) also indirectly affects the planetesimal formation process by altering the dust growth timescale. We also establish a control simulation  \texttt{run\_vert} based on  \texttt{run\_fid}. As can be seen from Table~\ref{tab3}, we reduce the value of \(\delta_{\rm z}\) from \(10^{-4}\) in \texttt{run\_fid} to \(10^{-6}\) in \texttt{run\_vert}. 
    This reduction decreases the dust scale height, which both increases the midplane dust-to-gas and shortens the growth timescale. However, $\epsilon$ is not high enough to trigger the SI at early times, and fast growth leads rapid depletion of dust mass. In the end we find that less dust is retained by the time that cavity opens, leading to a decrease in the final planetesimal mass from 31.4 \(M_\oplus\) to 20.9 \(M_\oplus\).

\subsection{Local pressure bump}
\label{Sect.4.3}
    Apart from the inner cavity generated by stellar X-ray photoevaporation, the pressure bumps associated with substructures can also be produced by other mechanisms, such as  gap opening by massive planets \citep{Kobayashi2012,zhu2012dust,gonzalez2015accumulation,eriksson2020pebble}, at the inner edge of dead zones \citep{varniere2006reviving,Lyra2009,Dzyurkevich2010}, by zonal flows arising from hydrodynamic instabilities \citep{lehmann2025convective} or magnetorotational turbulences \citep{johansen2009zonal}, and in the ice line regions \citep{ida2016formation,drkazkowska2017planetesimal,schoonenberg2017planetesimal}. The pressure bumps act as traps that halt the inward drift and concentrate dust particles. Under favorable conditions, this process can trigger the SI and lead to planetesimal formation.
    
    To further investigate the role of pressure bumps that may be produced by other mechanisms, we perform an additional simulation in which gaps are artificially imposed on the gas disk. The gap profile is adopted from \cite{Dullemond2018}
    \begin{equation}
        \alpha(r)=\alpha_0/F(r),
    \end{equation}
    where $\alpha_0$ is the disk viscosity, and $F(r)$ is given by
    \begin{equation}
        F(r)=\exp{\big[-A\exp{\left( -(r-r_0)^2/(2w^2) \right)}\big]},
    \end{equation}
    where $A$ is the gap amplitude, $r_0$ is the gap location, and $w$ is the gap width.

    Fig.~\ref{fig:bump} shows three snapshots of the dust continuum images at \(\rm 880\ \mu m\) with two artificial bumps at 30 au and 60 au, where $A{=}0.4$ and $w{=}0.1 r_0$. Other parameters are the same as \texttt{run\_fid}.
    At \(\rm t{=}0.38\ Myr\), the two wide gaps block the inward drifting of dust from the outer disk region before the onset of photoevaporative cavity (see Fig.~\ref{fig:bump}a). We can see that the photoevporative cavity forms and progessively moves outward. At \(\rm t{=}0.60\ Myr\) the cavity moves to 18 au (see Fig.~\ref{fig:bump}b), whereas at \(\rm t{=}0.80\ Myr\) it merges with the inner bump at 30 au (see Fig.~\ref{fig:bump}c).
    
    The bumps can also be sites for dust trapping and planetesimal formation. Consequently, the total planetesimal mass increases from $31.4\ M_\oplus$ to $33.9\ M_\oplus$. Meanwhile, ${\sim}7.6\ M_\oplus$ mass of planetesimals form in the bumps, leading to a slight reduction in the planetesimal mass formed at the outer edge of the photoevaporative cavity, from ${\sim}30.5\ M_\oplus$ to ${\sim}26.3\ M_\oplus$. We also note that as photoevaporative cavity expands progressively outward, it merges with the existing bumps.

    We compute the optical depth $\tau_{\nu}$ following 
    \begin{equation}
        \tau_{\nu} = \sum_i \left( \kappa_{\nu,i}^{\mathrm{abs}} + \kappa_{\nu,i}^{\mathrm{sca,eff}} \right) \Sigma_{\rm d}^i,
    \end{equation}
    where $\kappa_{\nu,i}^{\mathrm{abs}}$ and $\kappa_{\nu,i}^{\mathrm{sca,eff}} {=} (1 {-} \rm g_{\nu,i}) \kappa_{\nu,i}^{\mathrm{sca}}$ are the absorption and effective scattering opacities of dust species $i$, and $\Sigma_{\rm d}^i$ is the dust surface density of dust species $i$. The values of $\kappa_{\nu,i}^{\mathrm{abs}}$, $\kappa_{\nu,i}^{\mathrm{sca}}$ and $\rm g_{\nu,i}$ are adopted from DSHARP opacity model \citep{birnstiel2018disk}. The optical depth ($838\ \rm \mu m$) at the edge of photoevaporative cavity is ${\sim}2.1$ at $t{=}0.60\ \rm Myr$ (see Fig.~\ref{fig:opac}). The $\tau_{\nu}$ values of the three bumps are approximately in the ratio of 4:2.4:1, whereas the corresponding dust surface density ratios are 3:2:1. The relatively high optical depth at the cavity edge arises not only from the larger dust surface density but also from the preferential removal of small particles by photoevaporation, which increases the fraction of particles larger than 1 mm. Since such large particles have higher opacities at the observation wavelength, the resulting $\tau_{\nu}$ becomes significantly enhanced in this region.
    
    \cite{garate2023millimeter} also studied the combined impact of early substructures and photoevaporation. They found that in disks with primordial substructures, once photoevaporation opened an inner cavity, further dust loss due to radial drift was largely prevented, and dust removal by wind entrainment had only a minor effect on the evolution of solids. Such disks can retain a substantial dust mass.The difference between their work and ours is that they do not include the planetesimal formation. Furthermore, they indicated that the ratio of the gap location relative to the disk’s characteristic radius $r/R_{\rm c}$ controls how much dust can be retained.

\subsection{Caveats}
\label{Sect.4.4}
    We point out several limitations in our study. First, we neglect the interaction of dust onto gas, commonly referred to as dust-to-gas backreaction, in all of our simulations. This effect can be pronounced when the solid density approaches that of the surrounding gas ($\epsilon{=}\rho_{\mathrm{d}}/\rho_{\mathrm{g}}{\sim}1$). From the classical SI perspective, the drag force exerted by dust particles accelerates the gas's azimuthal motion. This, in turn, suppresses the inward drift of solids and further increases the local dust-to-gas ratio, making planetesimal formation more favorable.
    
    Nevertheless, near the pressure maximum of the photoevaporation cavity, this backreaction plays a dual role: whereas it slows down the drift of dust, the gas simultaneously gains angular momentum, which acts to widen the pressure bump. This broadening would on the other hand limit the further dust concentration. Consequently, the overall impact of backreaction on the SI and planetesimal formation can be more complicated. We also conduct an additional simulation based on \texttt{run\_fid} that includes the dust-to-gas backreaction, implemented through the \texttt{DustPyLib} interface \citep{stammler2023dustpylib}. The result is overall similar in terms of final planetesimal mass, which decreases from $31.4\ M_\oplus$ to $30.7 \ M_\oplus$.
    
    The other limitation is not accounting for the effect of the ice line.  Our model employs a static disk temperature profile and therefore does not account for the movement of the ice lines. Furthermore, we do not consider composition-dependent sublimation as pebbles drift across the ice lines of various species. The process of sublimation, followed by diffusion and re-condensation, could locally enhance the solid abundance near the ice line location \citep{ida2016formation,drkazkowska2017planetesimal,schoonenberg2017planetesimal}. Consequently, the omission of the above mechanisms likely leads to an underestimation of planetesimal formation near ice lines. Future models should incorporate an evolving disk thermal structure and track the compositional evolution of pebbles to properly model the influence of ice lines.

    Here we also do not consider the bouncing barrier. Laboratory and numerical studies show that the bouncing barrier typically prevents further coagulation once the dust aggregates reach mm/cm in size \citep{Zsom2010}. Neglecting this effect may therefore overestimate the mass fraction of pebbles that can grow to the sizes relevant for triggering the SI.

    Finally, we note that turbulence acts as a more complex role in realistic disk environments than that assumed in \cite{Lim_2024}. Large-scale structures developed from magnetorotational or hydrodynamic instabilities can create local pressure maxima that efficiently trap solids and promotes planetesimal formation. 
    In particular, \cite{eriksson2025} found that vortices can efficiently trap drifting dust at their pressure maxima, leading to locally enhanced dust-to-gas ratios. The weak turbulence and low relative velocities within the central regions of vortices facilitate particle growth, and once planetesimals form, they can continue to grow through efficient pebble accretion. Likewise, \cite{carrera2025positive} identified a positive feedback between dust coagulation and the SI, where enhanced dust concentration reduces turbulence and drift, promoting further particle growth and a stronger SI. Moreover, \cite{Huang_2025} demonstrated that the interplay among the VSI, the RWI and the SI can produce strong dust clumping within vortices in the outer disk regions even at moderately low dust-to-gas ratios. These effects are not considered in our current work.

\section{Conclusions}
\label{Sect.5}
In this work,  we employ the \texttt{DustPy} code to study the dust coagulation, fragmentation, and radial drift in a viscously accreting protoplanetary disk subject to stellar X-ray photoevaporation. Photoevaporation opens an inner cavity at a few au orbital distance, causing the disk's inner edge to expand radially outward as gas disperses. This process efficiently removes gas via photoevaporative winds, and therefore largely enhances the local dust-to-gas ratio at the pressure bump near the cavity's edge, which serves as a natural site to trigger the SI.  
We demonstrate that planetesimal formation via the SI, driven by the stellar X-ray photoevaporation, is robust during the disk dispersal phase. In our fiducial model, in total $31.4 \ M_{\oplus}$ of planetesimals form within the 100 au disk region. Furthermore, the exact choice of the SI criteria has little impact on the results.

We explore how the key physical processes regulate both the onset of the SI and the final masses of planetesimals.  We find that the efficiency of SI-induced planetesimal formation is strongly regulated by the available dust reservoir and its retention during disk evolution. 
Higher metallicity retains dust mass more effectively, directly enhancing planetesimal formation.  
Higher viscosity induces faster gas replenishment from the outer disk region, delays cavity opening, reduces overall dust reservoir, and thus hinders planetesimal formation. 
Higher stellar X-ray luminosity promotes earlier photoevaporative cavity opening while preserving dust at the cavity edge, thereby favoring planetesimal formation.
Larger disks prolong dust retention, thereby extending the duration of planetesimal formation. By retaining dust more efficiently, substructured disks create favorable conditions for planetesimal formation. 
To summarize, our models suggest that planetesimal formation is favored in disks with larger sizes, substructures, higher metallicities, lower viscosities, higher dust fragmentation threshold velocities, and/or around stars with higher X-ray luminosities.

Future work will focus on the growth and migration of these forming planetesimals \citep{fang2023planetesimal,lau2022rapid,lau2024sequential,lau2025formation} starting from a realistic birth mass distribution \citep{schafer2017initial,liu2020pebble}.

\bibliographystyle{aa}
\bibliography{reference}

\appendix
\section{Optical depth}
We compute the optical depth $\tau_{\nu}$ following 
\begin{equation}
    \tau_{\nu} = \sum_i \left( \kappa_{\nu,i}^{\mathrm{abs}} + \kappa_{\nu,i}^{\mathrm{sca,eff}} \right) \Sigma_{\rm d}^i,
\end{equation}
where $\kappa_{\nu,i}^{\mathrm{abs}}$ and $\kappa_{\nu,i}^{\mathrm{sca,eff}} {=} (1 {-} \rm g_{\nu,i}) \kappa_{\nu,i}^{\mathrm{sca}}$ are the absorption and effective scattering opacities of dust species $i$, and $\Sigma_{\rm d}^i$ is the dust surface density of dust species $i$. The values of $\kappa_{\nu,i}^{\mathrm{abs}}$, $\kappa_{\nu,i}^{\mathrm{sca}}$ and $\rm g_{\nu,i}$ are adopted from DSHARP opacity model \citep{birnstiel2018disk}. 

\begin{figure}[h]
    \centering
    \includegraphics[width=\linewidth]{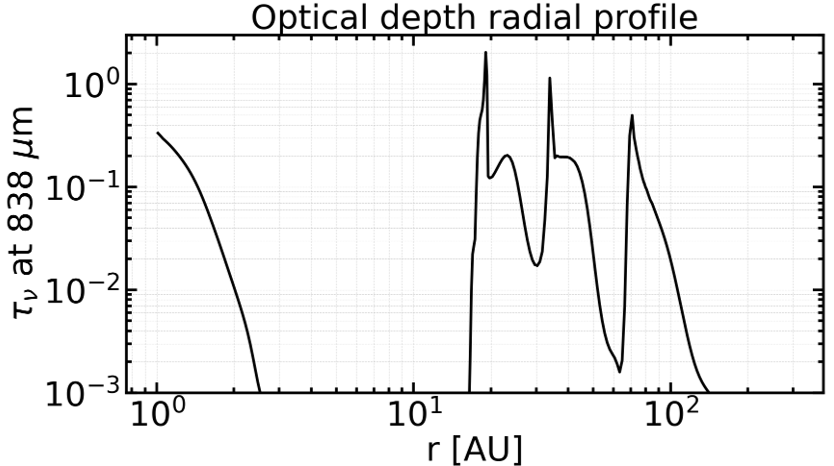}
    \caption{Radial profiles of the optical depth $\tau_{\nu}$ at $838\ \rm \mu m$ for the three pressure bumps at $t {=} 0.60\ \rm Myr$.}
    \label{fig:opac}
\end{figure}

The optical depth ($838\ \rm \mu m$) at the edge of photoevaporative cavity is ${\sim}2.1$ at $t{=}0.60\ \rm Myr$. The $\tau_{\nu}$ values of the other two bumps are approximately ${\sim} 1.2$ at ${\sim}34$ au and ${\sim}0.5$ at ${\sim} 70$ au. The optical depth of the outer bump at ${\sim} 70$ au is consistent with the results reported by \cite{Dullemond2018}, who showed that observed outer disk rings typically have optical depths of order ${\sim}0.5$. \cite{stammler2019dsharp} attributed these moderate optical depths to ongoing planetesimal formation, in agreement with our findings. In contrast, our results indicate significantly higher optical depths at the photoevaporative cavity edge and in the inner bump.

\section{Dust entrainment}
In this work, we adopt the dust entrainment model of \cite{sellek2020} as our fiducial prescription for dust removal by photoevaporation.
However, subsequent studies by \cite{booth2021modelling} and \cite{burn2022toward} suggested that the entrained dust size may be significantly smaller, motivating an alternative dust entrainment model driven by internal photoevaporation.
\begin{equation}
    a_{\rm ent,int} = \sqrt{\frac{8}{\pi}} 
    \frac{\dot{\Sigma}_{\rm{PE,g}}}{\rho_\bullet \Omega_{\rm K}}
    \frac{H_{\rm base}}{z_{\rm base}}
    \left(1 + \frac{z_{\rm{base}}^2}{r^2}\right)^{3/2},
\end{equation}
where $H_{\rm base} {\simeq} H_{\rm g}$ denotes the gas scale height at the wind base, and $z_{\rm base}$ is the height at which the photoevaporative wind is launched. We adopt $z_{\rm base}{/}H_{\rm base} {=} 4$ as suggested by \cite{booth2021modelling}. The maximum entrained size $a_{\rm ent,int}$ is limited to only a few microns.

\begin{figure}[h]
    \centering
    \includegraphics[width=\linewidth]{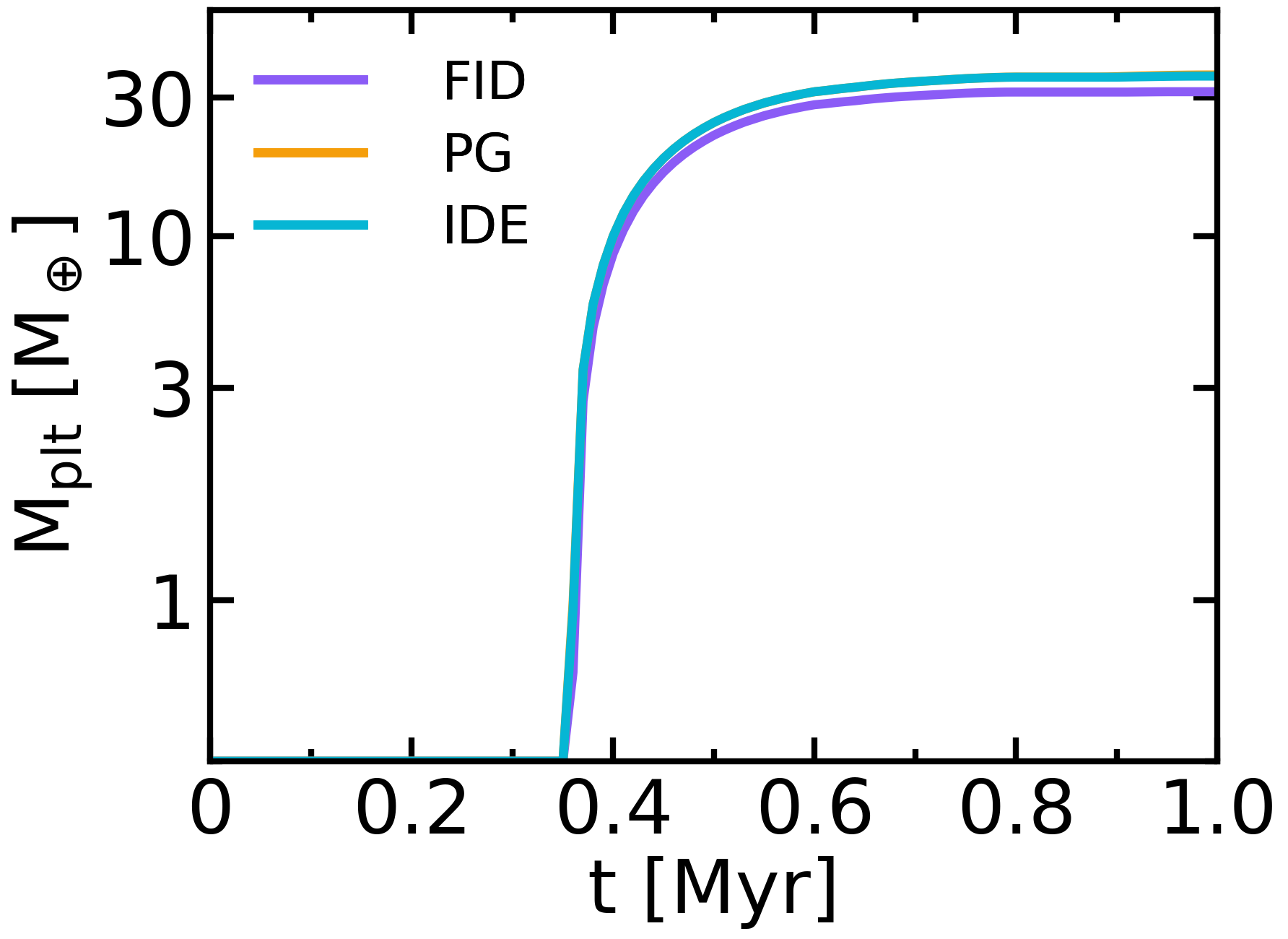}
    \caption{Time evolution of the planetesimal mass $M_{\rm plt}$ for three models FID, PG and IDE.}
    \label{fig:B1}
\end{figure}

\begin{figure}[h]
    \centering
    \includegraphics[width=\linewidth]{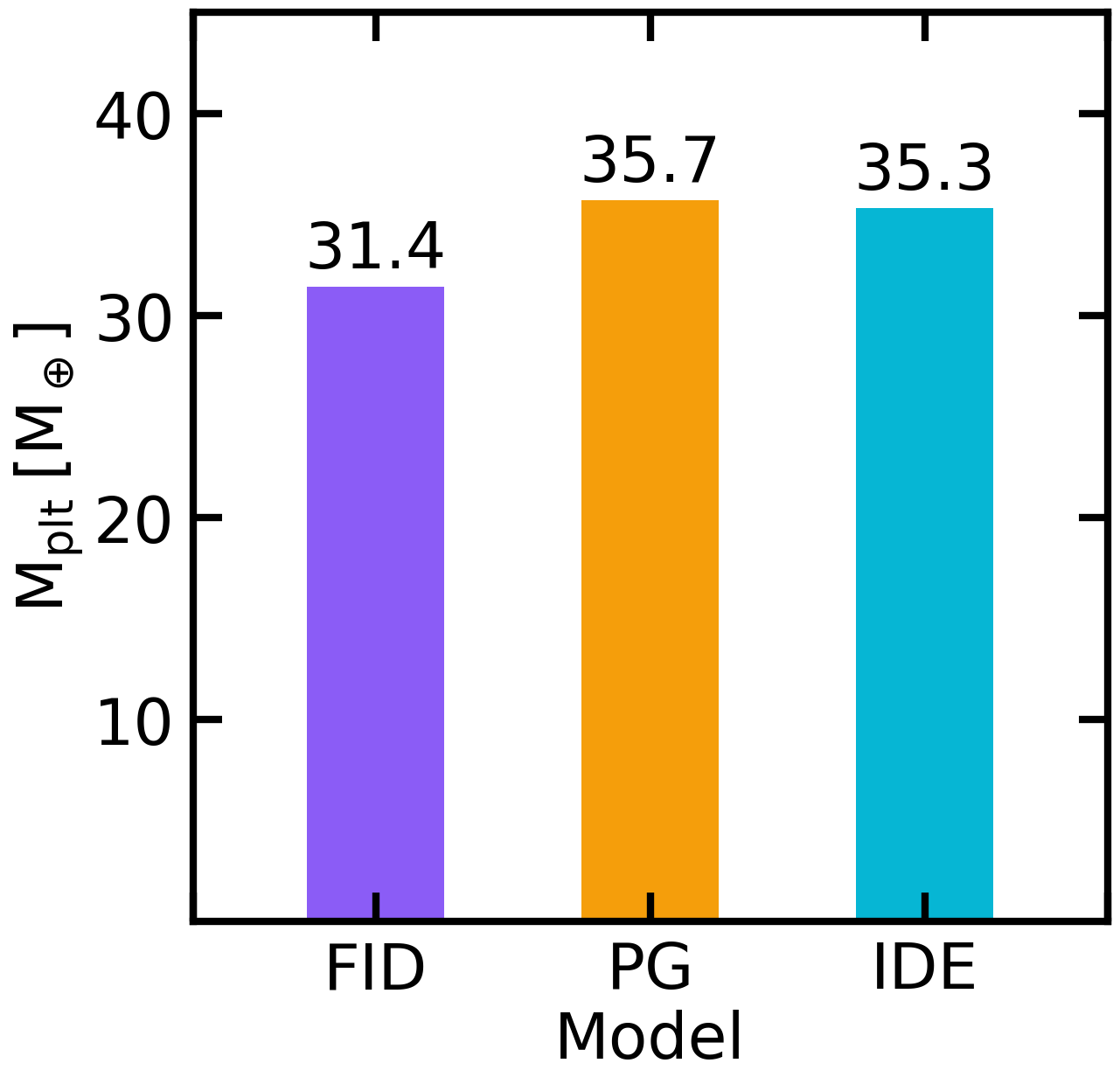}
    \caption{Total planetesimal mass $M_{\rm plt}$ for three models FID, PG and IDE.}
    \label{fig:B2}
\end{figure}

To assess the impact of these different assumptions, we perform a set of comparative simulations using three models: the fiducial model (FID), the pure gas photoevaporation model (PG), and the dust entrainment model driven by internal photoevaporation (IDE). Fig.~\ref{fig:B1} shows the time evolution of the total planetesimal mass $M_{\rm plt}$ for the three models, whereas Fig.~\ref{fig:B2} compares the final planetesimal mass obtained at the end of the simulations.

We find that there is no significant difference between the PG and IDE models, either in the time evolution of $M_{\rm plt}$ or in the final planetesimal mass.
This similarity arises because, in the IDE model, only dust particles with sizes of a few microns are efficiently entrained by the photoevaporative winds, leaving the bulk of dust mass largely unaffected.
In contrast, the FID model, which allows the removal of larger dust particles, systematically underestimates the final planetesimal mass. We will improve the treatment of dust entrainment in future work.

\end{document}